\documentclass[twocolumn]{aastex63}
\usepackage{float,graphicx,amsmath,multirow,mathtools}
\usepackage{color}
\usepackage[version=3]{mhchem}

\begin{document}

\title{The family of amide molecules toward NGC~6334I}

\correspondingauthor{Niels Ligterink}
\email{niels.ligterink@csh.unibe.ch}

\author[0000-0002-8385-9149]{Niels F.W. Ligterink}
\affiliation{Center for Space and Habitability (CSH), University of Bern, Sidlerstrasse 5, 3012 Bern, Switzerland}

\author{Samer J. El-Abd}
\affiliation{Department of Astronomy, University of Virginia, Charlottesville, VA 22904, USA}
\affiliation{National Radio Astronomy Observatory, Charlottesville, VA 22903, USA}

\author{Crystal L. Brogan}
\affiliation{National Radio Astronomy Observatory, Charlottesville, VA 22903, USA}

\author{Todd R. Hunter}
\affiliation{National Radio Astronomy Observatory, Charlottesville, VA 22903, USA}

\author{Anthony J. Remijan}
\affiliation{National Radio Astronomy Observatory, Charlottesville, VA 22903, USA}

\author{Robin T. Garrod}
\affiliation{Departments of Chemistry and Astronomy, University of Virginia, Charlottesville, VA 22904, USA}

\author{Brett M. McGuire}
\affiliation{National Radio Astronomy Observatory, Charlottesville, VA 22903, USA}
\affiliation{Harvard-Smithsonian Center for Astrophysics, Cambridge, MA 02138, USA}

\begin{abstract}
Amide molecules produced in space could play a key role in the formation of biomolecules on a young planetary object. However, the formation and chemical network of amide molecules in space is not well understood. In this work, ALMA observations are used to study a number of amide(-like) molecules toward the high-mass star-forming region NGC~6334I. The first detections of cyanamide (NH$_{2}$CN), acetamide (CH$_{3}$C(O)NH$_{2}$) and N-methylformamide (CH$_{3}$NHCHO) are presented for this source.  These are combined with analyses of isocyanic acid (HNCO) and formamide (NH$_{2}$CHO) and a tentative detection of urea (carbamide; NH$_{2}$C(O)NH$_{2}$). Abundance correlations show that most amides are likely formed in related reactions occurring in ices on interstellar dust grains in NGC~6334I. However, in an expanded sample of sources, large abundance variations are seen for NH$_{2}$CN that seem to depend on the source type, which suggests that the physical conditions within the source heavily influence the production of this species. The rich amide inventory of NGC~6334I strengthens the case that interstellar molecules can contribute to the emergence of biomolecules on planets.
\end{abstract}


\section{Introduction} \label{sec:intro}

\begin{figure}[ht!]
\includegraphics[width=0.5\textwidth]{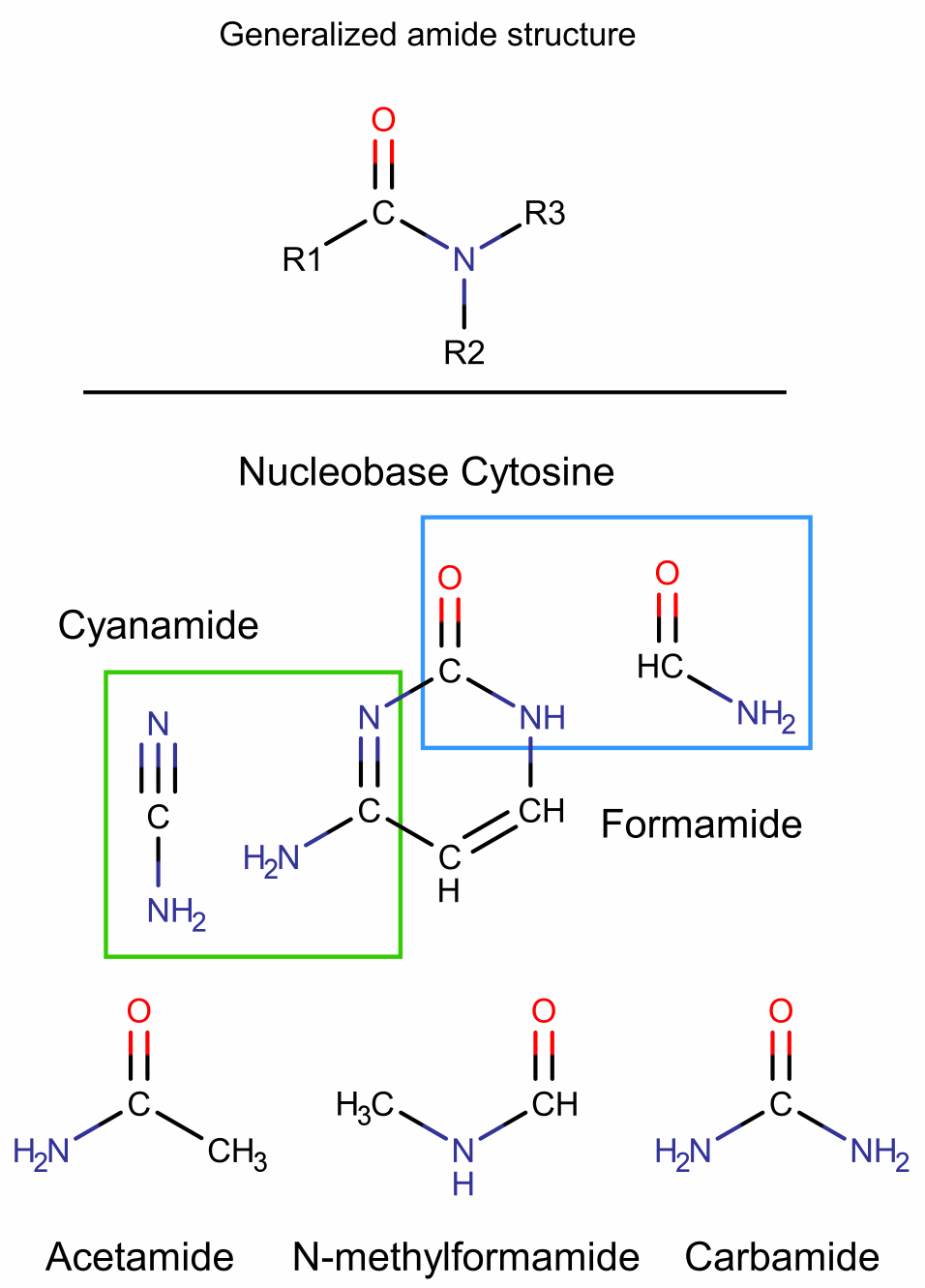}
\caption{Top: Generalized structure of an amide. Various functional group can be attached to the O=C-N backbone on locations R$_{1}$, R$_{2}$, and R$_{3}$ (R represents any arbitrary atom or functional group, not a specific element). Bottom: Molecular structures of amides observed in space and that of the nucleobase cytosine. Structural similarities between cytosine and NH$_{2}$CN and NH$_{2}$CHO are highlighted. \label{fig:func_groups}}
\end{figure}

Life as we know it relies on a select number of recurring chemical structures. Among these chemical structures, the amide unit (R$_{1}$-C(=O)N-R$_{2}$R$_{3}$, see top panel of Fig. \ref{fig:func_groups}), is an essential component in numerous biomolecules. Notable examples are nucleobases, the molecules that encode RNA and DNA, and as the link in the peptide chains that form proteins. In order to understand the chemical origin of life, it is important to understand how molecules with an amide unit came to be on Earth. Chemical reactions on a young Earth can give rise to this class of molecules \citep[e.g.,][]{patel2015}. However, a fascinating alternative is the formation of amides in space and subsequent delivery by, for example, meteorites to Earth \citep[e.g.,][]{chyba1990}. 

Several small amide molecules have been detected toward star and planet forming regions, most of them in recent years, such as formamide \citep[NH$_{2}$CHO,][]{rubin1971}, acetamide \citep[CH$_{3}$C(O)NH$_{2}$,][]{hollis2006}, and N-methylformamide \citep[CH$_{3}$NHCHO,][]{belloche2017,belloche2019}. The first evidence for carbamide, also known as urea (NH$_{2}$C(O)NH$_{2}$), in the gas phase in the ISM comes from \citet{remijan2014}, but it was securely identified by \citet{belloche2019}. Molecules that are closely related to the amide structure have been observed as well. Examples are cyanamide\footnote{Throughout this work, for convenience cyanamide will be referred to as an amide, although strictly speaking it does not follow the chemical definition of an amide. See appendix \ref{ap:naming} for an extended clarification on the naming and chemical structures of amides.} \citep[NH$_{2}$CN,][]{turner1975}, isocyanic acid \citep[HNCO,][]{snyder1972}, and methyl isocyanate \citep[CH$_{3}$NCO,][]{halfen2015,cernicharo2016,ligterink2017}. These molecules have structural similarities with biomolecules (see bottom panel of Fig. \ref{fig:func_groups}) and can act as precursor molecules to their formation.  

To assess the importance of interstellar amides for biochemistry, it is important to know if they commonly occur in star- and planet-forming regions. The simplest amide NH$_{2}$CHO and related molecule NH$_{2}$CN have been detected toward a number of inter- and circumstellar sources \citep{belloche2020}, including the sun-like protobinary IRAS~16293-2422 \citep[hereafter IRAS~16293]{kahane2013,coutens2016,coutens2018}.  NH$_{2}$CHO was also detected in comet 67P/Churyumov-Gerasimenko \citep[hereafter 67P/C-G]{goesmann2015,altwegg2017}. However, larger amides have almost exclusively been observed toward the galactic center source Sagittarius B2 \citep[hereafter Sgr B2,][]{halfen2011,cernicharo2016,ligterink2018a,belloche2017,belloche2019}. 

An expanded sample of amide detections can help elucidate both their interstellar chemistry and the interplay of that chemistry with physical processes during star-formation. For example, a large sample of HNCO and NH$_{2}$CHO observations clearly shows an abundance correlation between these two species over a variety of sources \citep[][and references therein]{lopez-sepulcre2019}. Hydrogenation (the addition of hydrogen atoms) of HNCO to form NH$_{2}$CHO has been proposed as a chemical explanation for the link between the two molecules \citep{lopez-sepulcre2015}. However, recent modeling work indicates that the relation between the two molecules is rather due to similar responses of the formation pathways of each molecule to the physical environment \citep{quenard2018}. Analogous analytical approaches could provide valuable insight into other amide species.  But while chemical reactions for such molecules have recently been included in chemical networks \citep{belloche2017,belloche2019}, observational constraints for these networks are sparse. 
To unravel interstellar amide chemistry and determine its relevance to biochemistry, it is important to expand the number of amide detections and to understand their formation. An ideal interstellar laboratory to study amides is found in the high-mass star-forming region (HMSFR) NGC~6334I. First, since the Solar System is thought to have formed in a massive cluster \citep[e.g.,][]{dukes2012}, NGC~6334I observations present a ``direct'' look at the conditions of early Solar System formation. Second, the chemistry of this region has been studied in great detail in the past decades, for example with the \emph{Herschel Space Observatory} \citep{zernickel2012}. In recent years, NGC~6334I has been extensively studied with the Atacama Large Millimeter Array (ALMA). This has not only resulted in a better understanding of the physical structure of this source \citep{brogan2016,brogan2018}, but also a better understanding of its molecular inventory. Various oxygen-bearing molecules have been investigated, including the C$_{2}$H$_{4}$O$_{2}$ isomers methyl formate (CH$_{3}$OCHO), glycolaldehyde (HC(O)CH$_{2}$OH), and acetic acid \citep[CH$_3$COOH,][]{mcguire2018,el-abd2019,xue2019}, deuterated methanol \citep{bogelund2018}, and even the first ever interstellar detection of methoxymethanol \citep[CH$_3$OCH$_2$OH,][]{mcguire2017} in presented toward this source. Additionally, a number of complex nitrogen-bearing molecules were investigated by \cite{bogelund2019a}, including NH$_{2}$CHO.

In this work, the inventory of amides toward NGC~6334I is investigated in detail. The first detections toward this object of NH$_{2}$CN, CH$_{3}$C(O)NH$_{2}$, and CH$_{3}$NHCHO are presented and combined with analysis of HNCO and NH$_{2}$CHO, and the tentative detection of NH$_{2}$C(O)NH$_{2}$. These observations and results are presented in section \ref{sec:methods} and \ref{sec:results}. The implications of the results are discussed in section \ref{sec:discussion}, before giving the conclusions in section \ref{sec:conclusion}. 

\begin{deluxetable*}{lccccccccc}
\tablecaption{Physical parameters for spectra analysed toward NGC~6334I \label{tab:source_params}}
\tablewidth{0pt}
\tablehead{
\colhead{Position} & \colhead{RA} & \colhead{DEC} &\colhead{$V_{\rm LSR}^{\dagger}$} & \colhead{$\Delta V$} & $T_{\rm ex}$ & \multicolumn{4}{c}{$T_{\rm BG}^{\ddagger}$} \\
& & & & & & (a) & (b) & (c) & (d) \\
\colhead{} & \colhead{hh:mm:ss} & \colhead{dd:mm:ss} & \colhead{(km s$^{-1}$)} & \colhead{(km s$^{-1}$)} & (K) & \multicolumn{4}{c}{(K)} 
}
\startdata
MM1-i & 17:20:53.373 & -35.46.58.341 & -7.0 & 3.25 & 135 & 10.1 & 26.9 & 31.3 & 31.3 \\
MM1-ii & 17:20:53.386 & -35.46.57.112 & -4.0 -- -4.5 & 3.25 & 175 & 38.9 & 79.7 & 94.6 & 85.1 \\
MM1-iii & 17:20:53.387 & -35.46.57.533 & -5.2 & 3.00 & 225 & 50.8 & 108.2 & 129.7 & 120.3 \\
MM1-iv & 17:20:53.420 & -35.46.59.088 & -8.2 & 3.50 & 150 & 15.7 & 38.4 & 44.8 & 73.6 \\
MM1-v & 17:20:53.434 & -35.46.57.856 & -4.4 & 3.25 & 285 & 69.0 & 159.3 & 192.7 & 176 \\
MM1-vi & 17:20:53.435 & -35.46.58.731 & -7.0 -- -7.5 & 3.25 & 190 & 30.2 & 77.6 & 88.6 & 94.4 \\
MM1-vii & 17:20:53.459 & -35.46.57.661 & -4.0 & 3.00 & 185 & 43.4 & 96.5 & 112.4 & 106.7 \\
MM1-viii & 17:20:53.469 & -35.46.58.724 & -6.8 & 3.00 & 150 & 25.8 & 57.1 & 64.7 & 81.9 \\
MM1-ix & 17:20:53.475 & -35.46.57.156 & -5.0 & 2.50 & 150 & 16.9 & 36.8 & 45.2 & 65.7 \\
MM1-nmf & 17:20:53.461 & -35.46.57.284 & -4.5 & 2.50 & 150 & 31.1 & 66.8 & 80.9 & 88.1 \\
MM2-i & 17:20:53.152 & -35.46.59.416 & -9.0 & 2.80 & 150 & 9.2 & 21.8 & 27.8 & 57.5 \\
MM2-ii & 17:20:53.178 & -35.46.59.494 & -9.0 & 2.80 & 200 & 17.4 & 44.4 & 58.6 & 98.4 \\
MM2-iii & 17:20:53.202 & -35.46.59.175 & -7.8 & 2.80 & 180 & 18.9 & 44.2 & 57.7 & 76.1 \\
\enddata
\tablecomments{$^{\dagger}$ For some positions certain molecules are fit with a different $V_{\rm LSR}$. For these cases a $V_{\rm LSR}$ range is given. $^{\ddagger}$Four different background temperature values are given for: (a) band 4, (b) band 7 $\leq$300 GHz, (c) band 7 $\geq$300 GHz, and (d) band 10 observations, respectively.}
\end{deluxetable*}


\section{Observations and analysis} \label{sec:methods}

\subsection{Observational Data of NGC~6334I}

\begin{figure*}[ht!]
\includegraphics[width=\textwidth]{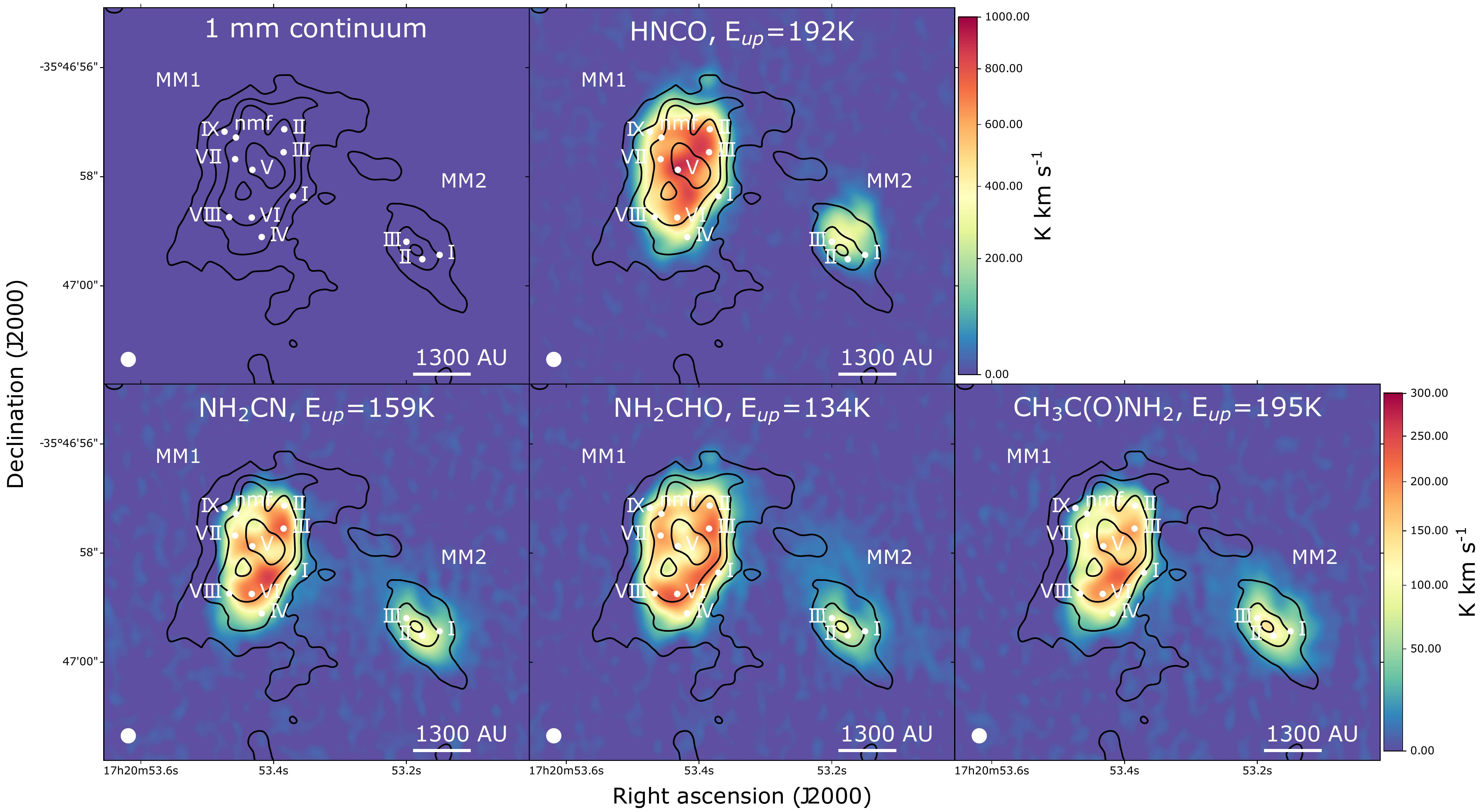}
\caption{Moment 0 maps of NH$_{2}$CN (transition: 14$_{2~13/12}$ - 13$_{2~12/11}$, 279.821 GHz, $E_{\rm up}$ = 159~K), HNCO (transition: 6$_{2~5/4}$ - 5$_{2~4/3}$, 131.847 GHz, $E_{\rm up}$ = 192~K), NH$_{2}$CHO (transition: 15$_{2~14}$ - 15$_{1~15}$, 281.935 GHz, $E_{\rm up}$ = 159~K), and CH$_{3}$C(O)NH$_{2}$ (transition: 25$_{3/2~23}$ - 24$_{2/3~22}$, 279.473 GHz, $E_{\rm up}$ = 195~K), with 1 mm dust continuum contours overplotted at 17.5, 52.5, 140, and 367.5 mJy beam$^{-1}$. The angular resolution of the observations (0.26\arcsec) is indicated by the bottom left circle. The positions of the extracted spectra are indicated by white dots labeled by capital roman numerals.  \label{fig:moment0}}
\end{figure*}

In this work, ALMA observations of NGC~6334I in bands 4, 7, and 10 are analyzed.  The details of the individual datasets and their reductions have been published in detail elsewhere \citep{mcguire2017,mcguire2018,brogan2018} and will only be discussed briefly here. The pertinent observing parameters for each set of observations are provided in Table~\ref{obsparams}. The data were self-calibrated and continuum-subtracted following the procedures detailed in \citet{hunter2017,brogan2018}. Prior to analysis, all of the observations were convolved to a common synthesized beam size of $0.26^{\prime\prime} \times 0.26^{\prime\prime}$. From the data cubes, background temperatures ($T_{\rm BG}$) were determined based on the continuum level temperature and spectra were extracted for analysis at the same positions as used in \citet{el-abd2019} and at an additional position called MM1-nmf. These positions are chosen to give an accurate overall sample of the chemistry toward NGC~6334I~MM1 and MM2. The parameters of these positions are listed in Table \ref{tab:source_params} and are indicated on moment 0 maps shown in Fig. \ref{fig:moment0}. In the case of Band 10 observations, robust analysis was only possible toward positions in MM1, as the MM2 source falls at the edge of the primary beam.

\begin{deluxetable*}{l c c c c}
\tablecaption{Summary of Pertinent ALMA Observing Parameters\label{obsparams}}
\tablewidth{0pt}
\tablehead{
\colhead{Parameter} & \colhead{Band 4} & \colhead{Band 7 Low} &\colhead{Band 7 High} & \colhead{Band 10}}
\startdata
Project Code                                &   \#2017.1.00661.S    &   \#2015.A.00022.T    &   \#2015.A.00022.T   &  \#2017.1.00717.S     \\
Configuration(s)                            &   C43-6               &   C36-4, C36-5        &   C36-4, C36-5        &   C43-3               \\
Primary Beam FWHM ($^{\prime\prime}$)            &   41                  &   20                  &   17                  &   6.6                 \\
Angular Resolution ($^{\prime\prime}$)$^{\dagger}$      &   $0.23\times0.16$    &   $0.25\times0.19$    &   $0.22\times0.17$    &   $0.21\times0.15$    \\
Frequency (GHz) & 130.03 -- 132.02 & 279.17 -- 282.94 & 336.18 -- 339.94 & 873.88 -- 881.33 \\
                & 144.00 -- 145.87 & 291.18 -- 294.94 & 348.18 -- 351.94 & 890.25 -- 897.70 \\
Spectral Resolution (km\,s$^{-1}$) & 1.1 & 1.1 & 1.1 & 0.5 \\
rms (mJy\,beam$^{-1}$\,km\,s$^{-1}$)        &   0.8                 &   2.0                 &   3.3                 &   44                  \\
\enddata
\tablecomments{$^{\dagger}$Note that all the observations were convolved to a common synthesized beam size of $0.26^{\prime\prime} \times 0.26^{\prime\prime}$.}
\end{deluxetable*}


\subsection{Molecule identification \& analysis} \label{sec:molid}
All spectra are analyzed with the CASSIS\footnote{CASSIS has been developed by IRAP-UPS/CNRS (http://cassis.irap.omp.eu).} line analysis software. Line lists from the JPL catalog for molecular spectroscopy \citep{pickett1998}, the Cologne Database for Molecular Spectroscopy \citep[CDMS,][]{muller2001,muller2005}, and literature are used. For the molecules of interest to this work, that is amide and amide-like molecules, an overview of database and literature sources is given in Table \ref{tab:spec_lit} in Appendix \ref{sec:spec_used}.

Given a line list and input parameters such as column density ($N_{\rm s}$), excitation temperature ($T_{\rm ex}$), source velocity ($V_{\rm LSR}$), and line width at full width at half maximum ($\Delta$V), CASSIS is able to generate synthetic molecular spectra. Molecular line emission is coupled with the background temperature. Normally, this is the cosmic microwave background temperature of 2.7~K, which has a negligible influence on molecular emission. However, background dust continuum temperatures are significantly higher toward NGC~6334I and it is therefore important to take this effect into account to properly simulate molecular spectra. $T_{\rm BG}$ varies for every position and frequency range toward NGC~6334I. Its values are given in Table \ref{tab:source_params}. Due to the extended molecular emission in NGC~6334I and the relatively small beam size, the emission is assumed to be beam filling and thus has a beam filling factor of one. 

The analysis of spectral lines is performed under the assumption that molecules are in local thermodynamic equilibrium (LTE) and have a single excitation temperature. The excitation temperature is determined from the observed line brightness temperature ($\Delta$ T$_{\rm B}$) above the background continuum of optically thick rotational lines and $T_{\rm BG}$ following the formalism of \citet{turner1991} \citep[see also section 2.2.1. of][]{el-abd2019}. Toward each positions all molecules are thus analyzed with the same excitation temperature. In the first step of the data analysis, rotational lines of a molecule are identified and a by-eye synthetic fit of these lines is made. From the by-eye fit the $V_{\rm LSR}$ and $\Delta$V are determined, which in most cases yield similar values as given in \cite{el-abd2019}. The errors on both the $V_{\rm LSR}$ and $\Delta$V are conservatively estimated to be $\pm$0.3~km~s$^{-1}$. Then, by cross checking with the spectral line databases, optically thin lines ($\tau \ll$ 1) are selected that are free of line blends with other species, although blends in the line wings of other species may occur. CASSIS then performs a regular grid $\chi^2$ minimization routine on the selected lines, where the column density is given as a free parameter within $\pm$1 order of magnitude of the by-eye determined value. The best fit (i.e., lowest $\chi^{2}$ value) determines the column density. In most cases, the error on the column density is given by the 1$\sigma$ fit error plus a 10\% flux uncertainty.A larger error of 30\% is used for for CH$_{3}$NHCHO, due to higher level of line blending. Further details on the CASSIS line analysis procedure can be found in \citet{ligterink2018b} and \citet{bogelund2019a}.

\section{Results} \label{sec:results}

\begin{figure*}[ht!]
\includegraphics[width=\textwidth]{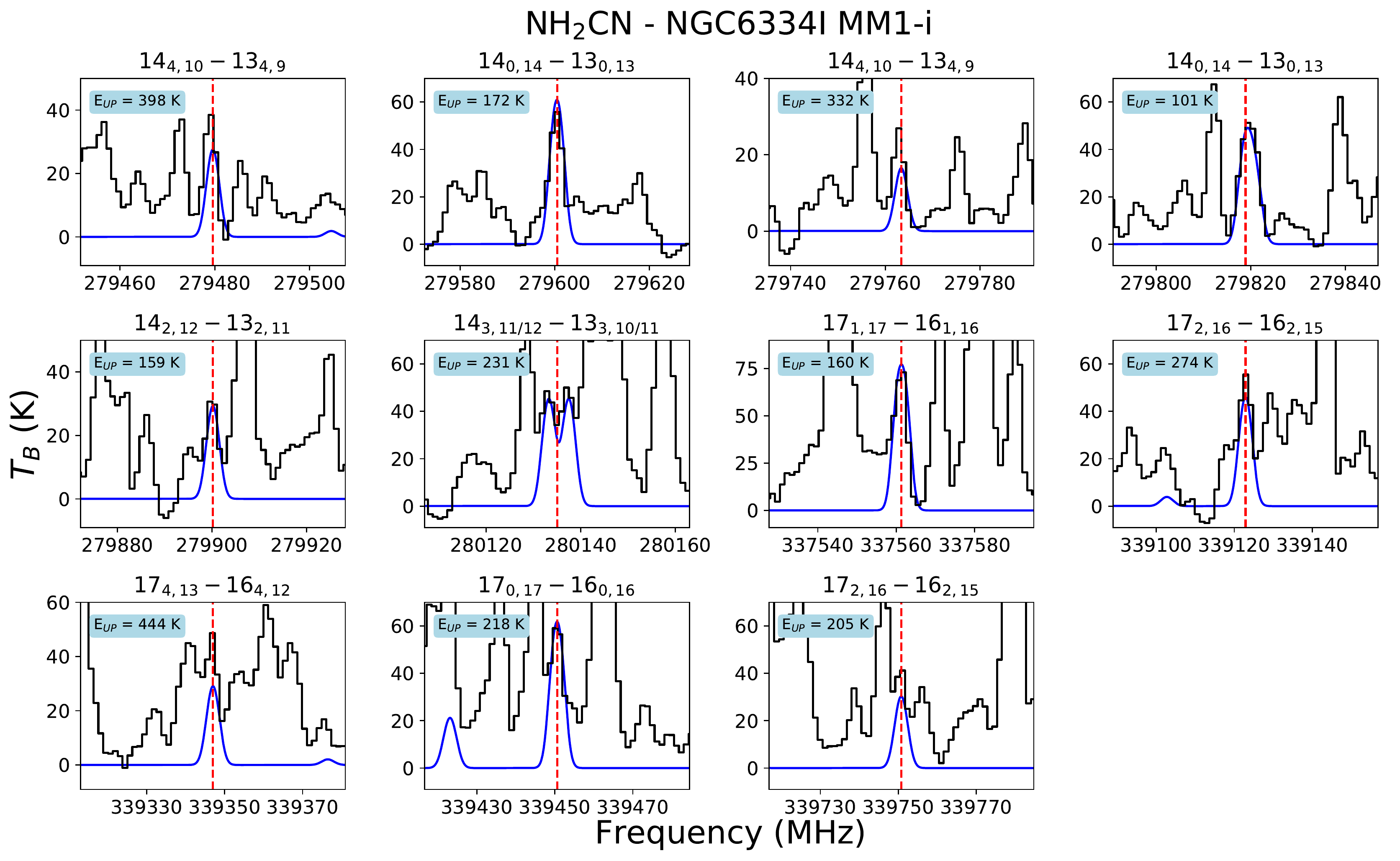}
\caption{Detected transitions of NH$_{2}$CN in the spectrum toward position MM1-i. The observed spectrum is plotted in black, with the synthetic spectrum overplotted in blue and the rest frequency center of the transition is indicated by the red dotted line. The upper state energy of each transition is given in the top left corner. \label{fig:cyanamide_mm1-i}}
\end{figure*}

\begin{figure*}[ht!]
\includegraphics[width=\textwidth]{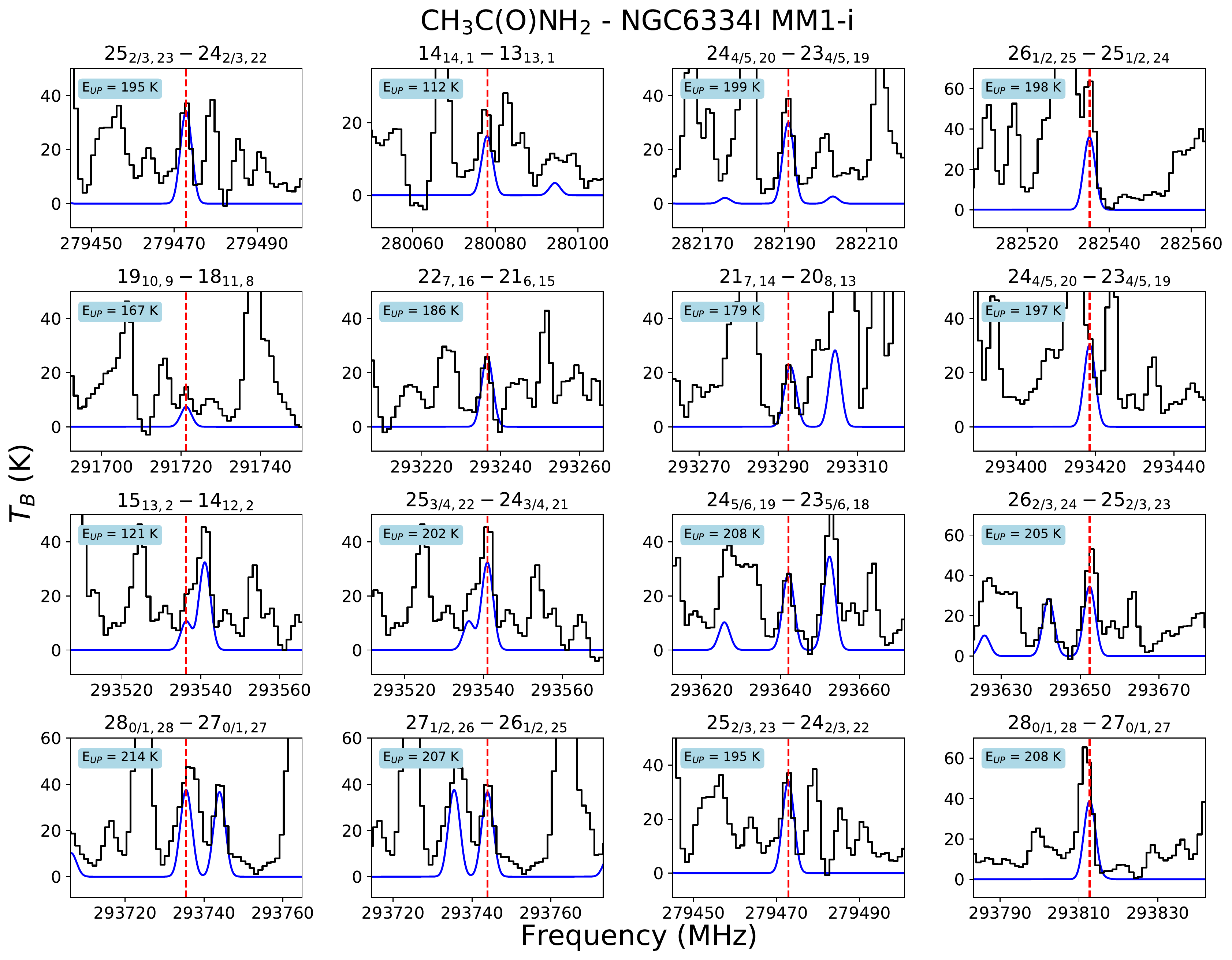}
\caption{Detected transitions of CH$_{3}$C(O)NH$_{2}$ in the spectrum toward position MM1-i. The observed spectrum is plotted in black, with the synthetic spectrum overplotted in blue and the rest frequency center of the transition is indicated by the red dotted line. The upper state energy of each transition is given in the top left corner. \label{fig:acetamide_mm1-i}}
\end{figure*}

Except for position MM2-iii, molecular line emission of NH$_{2}$CN, HN$^{12}$CO, HN$^{13}$CO, NH$_{2}^{12}$CHO, NH$_{2}^{13}$CHO, and CH$_{3}$C(O)NH$_{2}$ are detected toward all positions. The detected lines are listed in Table \ref{tab:lines_amides}. Many of the lines of the main isotopologues of HNCO and NH$_{2}$CHO are optically thick and therefore not used for further analysis. Following the routine described in section \ref{sec:molid}, the spectra of the remaining molecules were simulated and column densities derived. NH$_{2}$CN and CH$_{3}$C(O)NH$_{2}$ are detected for the first time toward NGC~6334I and their detected spectral lines toward position MM1-i are shown in Figs. \ref{fig:cyanamide_mm1-i} and \ref{fig:acetamide_mm1-i}. The spectra of the other molecules toward MM1-i and all detections toward MM2-i are presented in Appendix \ref{sec:mm1-2_spec_feat}. The derived column densities toward each position are listed in Table \ref{tab:coldens}. For further analysis, the column densities of HN$^{12}$CO and NH$_{2}^{12}$CHO are determined from their optically thin $^{13}$C counterparts. To derive the main isotopologue column density, the column densities of the $^{13}$C isotopologues are multiplied with the $^{12/13}$C ratio of 62 for NGC~6334I, as determined by \citet{bogelund2018}, based on equations given by \citet{milam2005}. Moment 0 maps of lines of NH$_{2}$CN, HN$^{12}$CO, NH$_{2}^{12}$CHO, and CH$_{3}$C(O)NH$_{2}$ overplotted on dust continuum contours are shown in Fig. \ref{fig:moment0}.

Toward all positions, spectral features are identified that can be assigned to CH$_{3}$NHCHO, although many of these lines are blended in the wings of other spectral features or on locations where the baseline dips. Despite this, a number of unblended CH$_{3}$NHCHO lines can be identified in the spectra of positions MM1-v, vi, vii, and viii and in particular MM1-nmf, which makes it possible to claim the first detection of CH$_{3}$NHCHO toward NGC~6334I. The detected CH$_{3}$NHCHO lines toward position MM1-nmf are shown in Fig. \ref{fig:nmf_mm1-nmf}. Figure \ref{fig:nmf_mm1-nmf_dips} in Appendix \ref{sec:mm1-2_spec_feat} show the CH$_{3}$NHCHO lines in the MM1-nmf spectrum that are present on locations where the baseline dips and thus are not well reproduced by the synthetic spectrum. The spectral features identified toward the other positions in MM1 and in MM2 are insufficient to claim a detection and therefore CH$_{3}$NHCHO is tentatively identified toward these positions. Column densities of this species are presented in Table \ref{tab:coldens}.

\begin{figure*}[ht!]
\includegraphics[width=\textwidth]{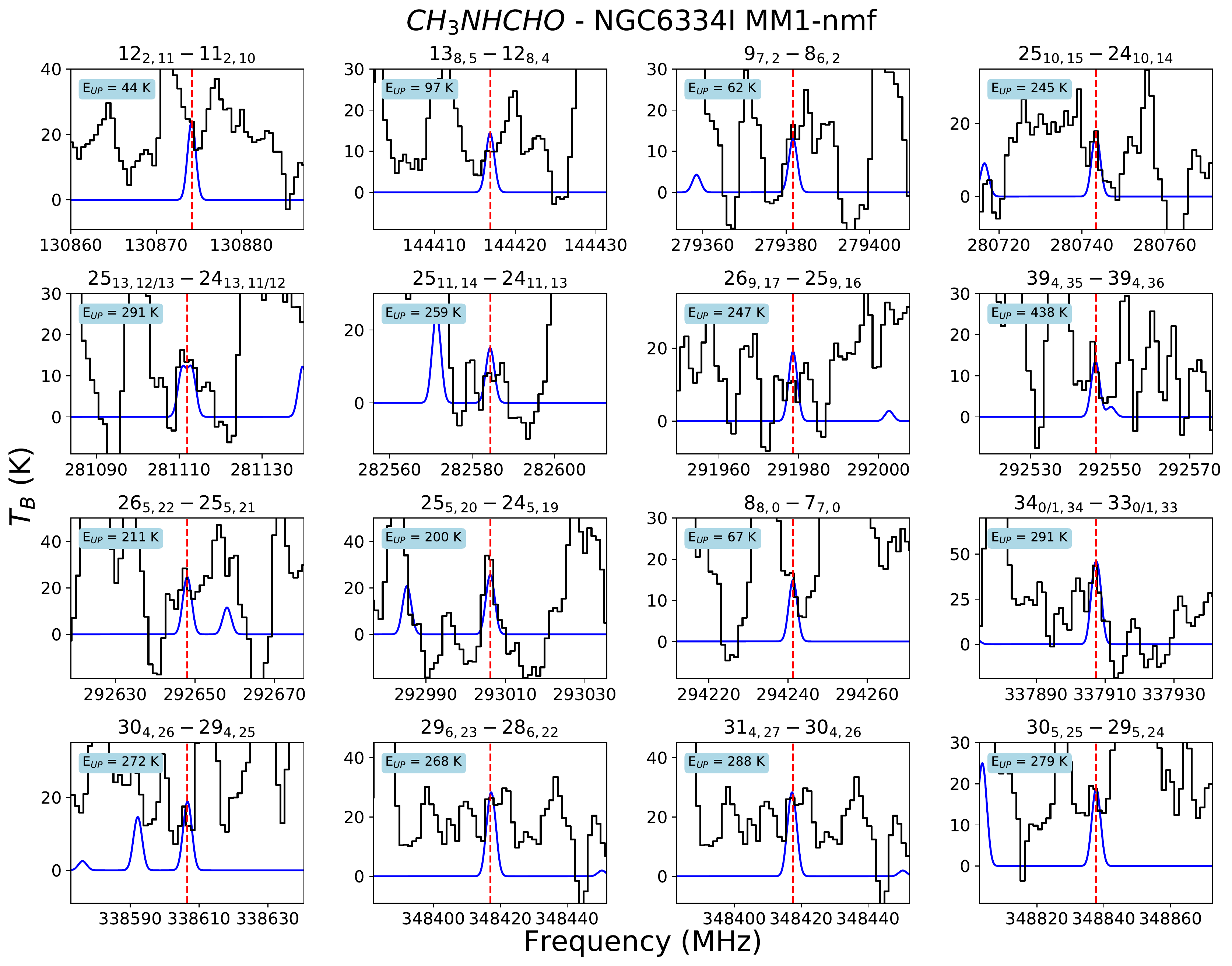}
\caption{Detected transitions of CH$_{3}$NHCHO in the spectrum toward position MM1-nmf. The observed spectrum is plotted in black, with the synthetic spectrum overplotted in blue and the rest frequency center of the transition is indicated by the red dotted line. The upper state energy of each transition is given in the top left corner.  \label{fig:nmf_mm1-nmf}}
\end{figure*}

\begin{figure*}[ht!]
\includegraphics[width=\textwidth]{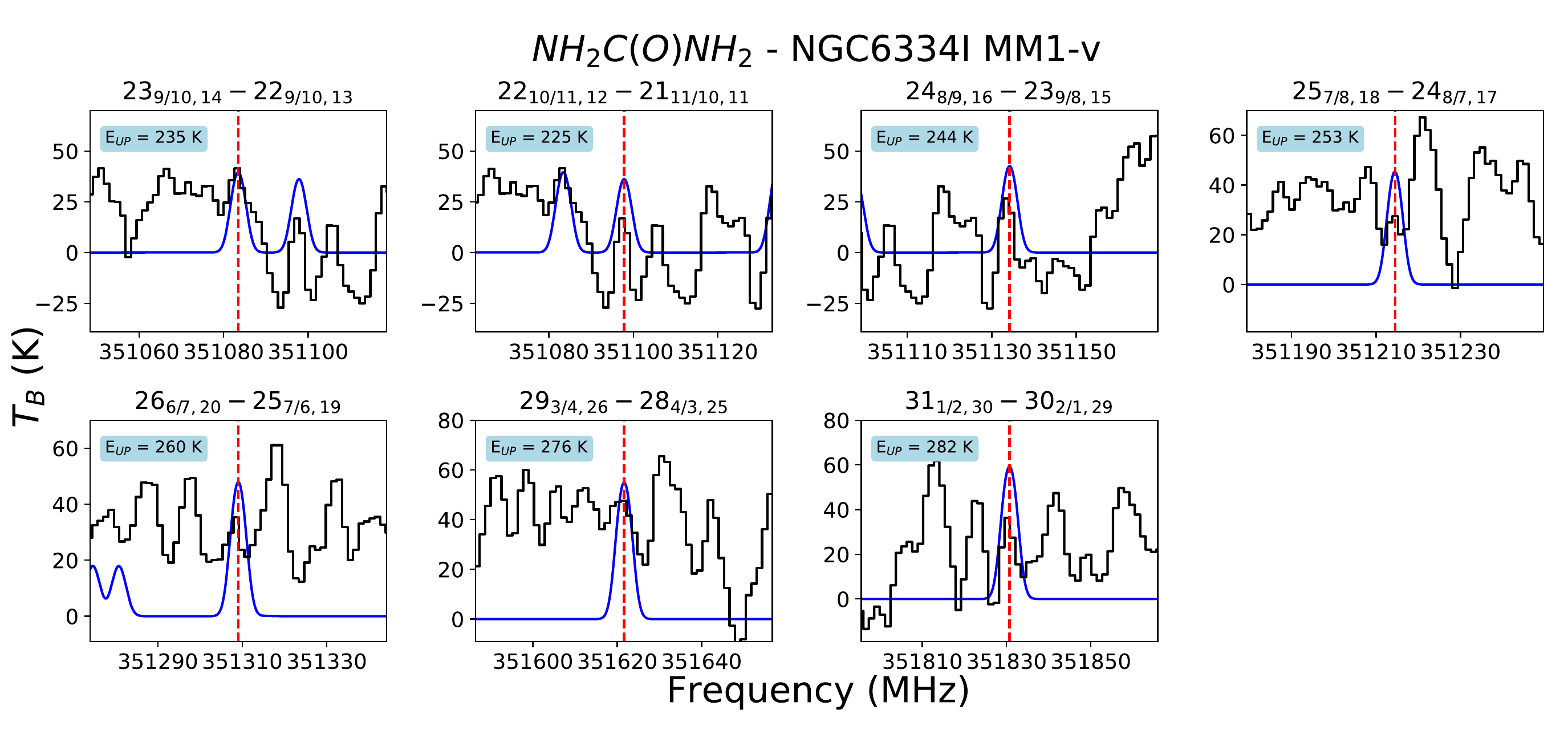}
\caption{Overview of NH$_{2}$C(O)NH$_{2}$ spectral features toward position MM1-v. The observed spectrum is plotted in black, with the synthetic spectrum overplotted in blue and the rest frequency center of the transition is indicated by the red dotted line. The upper state energy of each transition is given in the top left corner. \label{fig:urea_mm1-v}}
\end{figure*}

At a number of positions a handful of spectral features are identified that can be assigned to NH$_{2}$C(O)NH$_{2}$. The most and clearest lines are identified toward position MM1-v, see Fig. \ref{fig:urea_mm1-v}, with at least three unblended lines. While these lines are not enough for a secure detection, a tentative identification can be claimed. The tentative and upper limit column densities of NH$_{2}$C(O)NH$_{2}$ are presented in Table \ref{tab:coldens}.

\begin{deluxetable*}{lccccccccc}
\movetabledown=100mm
\tablecaption{Column densities and upper limits of amides analysed in this work \label{tab:coldens}}
\tablewidth{0pt}
\tablehead{
\colhead{Position} & \colhead{$T_{\rm ex}$} & \colhead{$N$(NH$_{2}$CN)} & \colhead{$N$(HN$^{13}$CO)} &  \colhead{$N$(NH$_{2}^{13}$CHO)} & \colhead{$N$(CH$_{3}$C(O)NH$_{2}$)} & \colhead{$N$(CH$_{3}$NHCHO)} & \colhead{$N$(NH$_{2}$C(O)NH$_{2}$)} \\
\colhead{} & \colhead{(K)} & \colhead{cm$^{-2}$} & \colhead{cm$^{-2}$} & \colhead{cm$^{-2}$} & \colhead{cm$^{-2}$} & \colhead{cm$^{-2}$} & \colhead{cm$^{-2}$} 
}
\startdata
MM1-i & 135 & (5.9$\pm$0.9)$\times$10$^{15}$ & (9.9$\pm$3.0)$\times$10$^{15}$ & (8.6$\pm$1.4)$\times$10$^{15}$ &  (4.7$\pm$0.7)$\times$10$^{16}$ & $\lnapprox$(7.5$\pm$2.8)$\times$10$^{16}$ & $\leq$2.0$\times$10$^{15}$ \\
MM1-ii & 175 &  (1.8$\pm$0.3)$\times$10$^{16}$ & (4.5$\pm$1.4)$\times$10$^{16}$ & (2.2$\pm$0.6)$\times$10$^{16}$ & (7.4$\pm$1.1)$\times$10$^{16}$ & $\leq$1.5$\times$10$^{17}$ & $\leq$5.0$\times$10$^{15}$ \\
MM1-iii & 225 &  (3.5$\pm$0.5)$\times$10$^{16}$ & (6.0$\pm$1.8)$\times$10$^{16}$ & (3.2$\pm$0.5)$\times$10$^{16}$ & (1.7$\pm$0.4)$\times$10$^{17}$ & $\leq$2.5$\times$10$^{17}$ & $\leq$1.0$\times$10$^{16}$   \\
MM1-iv & 150 & (7.8$\pm$1.6)$\times$10$^{15}$ & (1.3$\pm$0.2)$\times$10$^{16}$ & (1.3$\pm$0.2)$\times$10$^{16}$ & (4.0$\pm$1.2)$\times$10$^{17}$ & $\lnapprox$(1.0$\pm$0.4)$\times$10$^{17}$ & $\leq$1.0$\times$10$^{16}$  \\
MM1-v & 285 & (4.0$\pm$0.4)$\times$10$^{16}$ & (1.2$\pm$0.1)$\times$10$^{17}$ & (2.7$\pm$0.3)$\times$10$^{16}$ & (2.3$\pm$0.2)$\times$10$^{17}$ & (3.6$\pm$0.7)$\times$10$^{17}$ & $\lnapprox$(5.0$\pm$2.0)$\times$10$^{16}$  \\
MM1-vi & 190 & (2.6$\pm$0.6)$\times$10$^{16}$ & (3.7$\pm$0.5)$\times$10$^{16}$ & (2.8$\pm$0.3)$\times$10$^{16}$ & (1.6$\pm$0.2)$\times$10$^{17}$ & (2.0$\pm$0.6)$\times$10$^{17}$ & $\leq$1.0$\times$10$^{16}$  \\
MM1-vii & 185 & (1.6$\pm$0.3)$\times$10$^{16}$ & (4.4$\pm$0.5)$\times$10$^{16}$ & (3.8$\pm$0.4)$\times$10$^{16}$ & (1.6$\pm$0.2)$\times$10$^{17}$ & (2.0$\pm$0.6)$\times$10$^{17}$ & $\leq$3.0$\times$10$^{16}$  \\
MM1-viii & 150 & (7.8$\pm$0.8)$\times$10$^{16}$ & (2.0$\pm$0.3)$\times$10$^{16}$ & (1.6$\pm$0.3)$\times$10$^{16}$ & (7.1$\pm$1.1)$\times$10$^{16}$ & (1.4$\pm$0.5)$\times$10$^{17}$ & $\leq$2.0$\times$10$^{16}$  \\
MM1-ix & 150 & (2.3$\pm$0.4)$\times$10$^{15}$ & (6.4$\pm$1.3)$\times$10$^{15}$ & (4.8$\pm$0.5)$\times$10$^{15}$ & (1.6$\pm$0.3)$\times$10$^{16}$ & $\lnapprox$(4.8$\pm$1.5)$\times$10$^{16}$ & $\leq$1.5$\times$10$^{16}$  \\
MM1-nmf & 150 & (1.1$\pm$0.2)$\times$10$^{16}$ & (1.9$\pm$0.3)$\times$10$^{16}$ & (1.6$\pm$0.3)$\times$10$^{16}$ & (6.8$\pm$1.3)$\times$10$^{16}$ & (1.0$\pm$0.5)$\times$10$^{17}$ & $\leq$1.0$\times$10$^{16}$ \\
MM2-i & 150 & (3.5$\pm$0.6)$\times$10$^{15}$ & (1.4$\pm$0.5)$\times$10$^{16}$ & (3.3$\pm$1.0)$\times$10$^{15}$ & (5.9$\pm$1.6)$\times$10$^{16}$ & $\leq$5.0$\times$10$^{16}$ & $\leq$3.0$\times$10$^{15}$  \\
MM2-ii & 200 &  (4.0$\pm$0.5)$\times$10$^{15}$ & (2.3$\pm$0.7)$\times$10$^{16}$ & (3.8$\pm$1.1)$\times$10$^{15}$ & (8.8$\pm$1.6)$\times$10$^{16}$ & $\leq$6.0$\times$10$^{16}$ & $\leq$4.0$\times$10$^{15}$  \\
\enddata
\end{deluxetable*}


\section{Discussion} \label{sec:discussion}

\begin{figure*}[ht!]
\includegraphics[width=\textwidth]{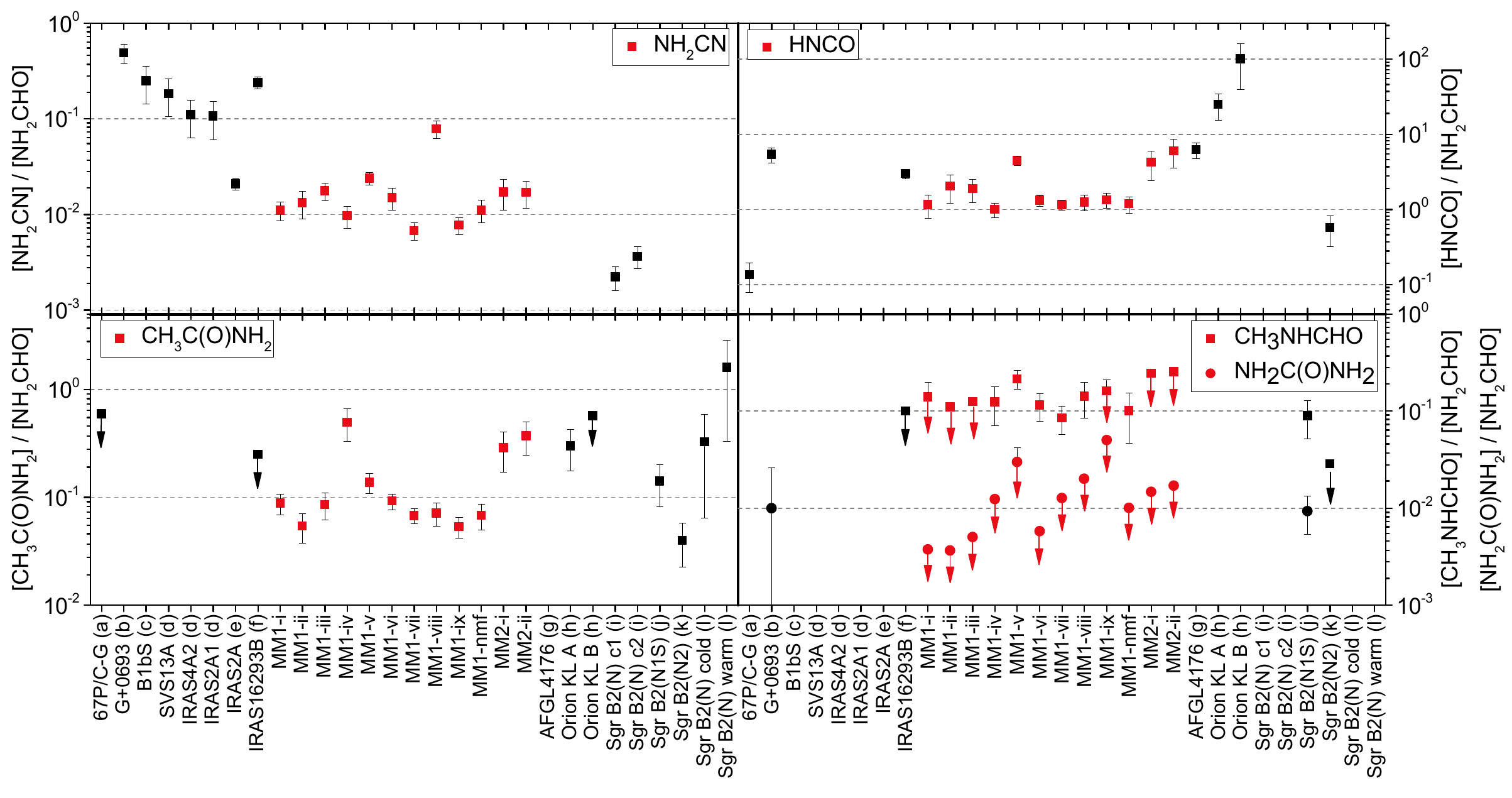}
\caption{Column density ratios of NH$_{2}$CN (top left), HNCO (top right), CH$_{3}$C(O)NH$_{2}$ (bottom left), and CH$_{3}$NHCHO and NH$_{2}$C(O)NH$_{2}$ (bottom right) with respect to NH$_{2}$CHO. Ratios for NGC~6334I~MM1 and MM2 are indicated in red. For the NGC~6334I data, $^{13}$C isotopes of HNCO and NH$_{2}$CHO are used, multiplied by the $^{12/13}$C ratio of 62. The sources are placed in the order: comet 67P, quiescent  molecular cloud, low-mass protostars / hot corinos, NGC~6334I, and other HMSFRs, high-mass protostars, and hot cores. Data are taken from: $^{a}$\citet{goesmann2015,altwegg2017}, $^{b}$\citet{zeng2018,jimenez-serra2020}, $^{c}$\citet{marcelino2018}, $^{d}$\citet{belloche2020}, $^{e}$\citet{coutens2018}, $^{f}$\citet{coutens2016,coutens2018,ligterink2018a}, $^{g}$\citet{bogelund2019b}, $^{h}$\citet{cernicharo2016}, $^{i}$\citet{belloche2013}, $^{j}$\citet{belloche2017}, $^{k}$\citet{belloche2019}, and $^{l}$\citet{hollis2006}.  \label{fig:ratios_formamide}}
\end{figure*}

\subsection{Chemical comparisons}
To get a better understanding of the chemical processes leading to amides, it is useful to compare molecule ratios of the positions within NGC~6334I, but also with other inter- and circumstellar sources. In the following sections molecular ratios with respect to NH$_{2}$CHO are discussed. Comparisons are made with the following works: IRAM~30m and Green Bank Telescope (GBT) single dish observations of G+0.693-0.027 (hereafter G+0.693), a quiescent giant molecular cloud \citep{zeng2018,jimenez-serra2020}, ALMA observations of the incipient hot corino Barnard B1b-S \citep{marcelino2018}, Northern Extended Millimeter Array (NOEMA) observations of the low-mass protostars SVS13A, NGC~1333-IRAS4A2, NGC~1333-IRAS2A(1) \citep{coutens2018,belloche2020}, ALMA Protostellar Interferometric Line Survey \citep[PILS,][]{jorgensen2016} observations of the low-mass protostar IRAS~16293B \citep{coutens2016,coutens2018,ligterink2018a}, GBT and IRAM 30m single dish observations of the galactic center source Sgr B2 \citep{hollis2006,belloche2013}, ALMA observations of the HMSFR Orion KL \citep{cernicharo2016}, the high-mass star AFGL 4176 \citep{bogelund2019b}, and of Sgr B2 \citep{belloche2017,belloche2019}. One has to be aware that precise knowledge of the source size or emitting area of a certain molecule is essential for ratio comparisons. In particular, single dish telescopes do not have the spatial resolution to determine the source size and these are often assumed or estimated, potentially resulting in large column density errors. Finally, mass spectrometric measurements of the Rosetta mission to comet 67P/C-G are used to compare with a Solar System object \citep{goesmann2015,altwegg2017}. The sources are placed in the order: comet 67P, quiescent  molecular cloud, low-mass protostars / hot corinos, NGC~6334I, and other HMSFRs, high-mass protostars, and hot cores. The derived ratios are presented in Fig. \ref{fig:ratios_formamide}. 


\subsubsection{Variations in NH$_{2}$CN abundance}

Within NGC~6334I, the NH$_{2}$CN over NH$_{2}$CHO ratio shows only minor variations and in general is uniform at $\sim$1-2\%. More noticeable is the variation between other sources, which span almost three orders of magnitude. Interestingly, the highest ratios are found for the quiescent molecular cloud and low-mass protostars, while lower ratios are seen in the high-mass sources NGC~6334I and Sgr B2. An important deviation in this pattern is seen in Orion KL, where [NH$_{2}$CN]/[NH$_{2}$CHO] = 0.4 -- 1.4 \citep[][data point not included in Fig. \ref{fig:ratios_formamide}]{white2003}. However, it is important to note that the column density of NH$_{2}$CHO in Orion KL was determined using a single spectral line and there is some uncertainty in the assumed excitation temperatures. If this trend is in fact dependent on the source type, it indicates that the source influences either the formation pathway or desorption from grain surfaces of one of the two species. 

Recently, \citet{coutens2018} used chemical models to explain the presence of NH$_{2}$CN toward IRAS 16293-2422B. Chemical networks for this molecule are very sparse, and no plausible gas phase mechanism for its production is currently evident. \citet{coutens2018} introduced a formation route for NH$_{2}$CN on grain surfaces through the reaction between NH$_{2}$ and CN radicals. In their chemical models they determined that this reaction is sensitive to the gas phase H and H$_{2}$ density: at high gas densities ($>10^9$ cm$^{-3}$), hydrogenation of the NH$_{2}$ and CN radicals on the ice mantle surface becomes important and the production of NH$_{2}$CN stagnates. Since HMSFRs generally speaking have higher gas densities, this could explain the trend in the observational data if particular sources enter this density regime for long periods when NH$_{2}$CN might otherwise be forming.

The formation of NH$_{2}$CHO is still a topic of debate, with proponents of both gas phase and solid state ice mantle formation routes \citep[see][for an overview of this discussion]{lopez-sepulcre2019}. Several computational studies have shown that NH$_{2}$CHO can be formed in the gas phase by a near barrierless reaction between NH$_{2}$ and H$_{2}$CO \citep[e.g.,][]{barone2015,skouteris2017}:
\begin{equation}
    \ce{NH2 + H2CO -> NH2CHO + H}.
\end{equation}
Many studies have been dedicated to understanding the formation of NH$_{2}$CHO in ice mantles, with the main proposed formation routes being the radical reaction between NH$_{2}$ and CHO directly yielding \ce{NH2CHO},
\begin{equation}
    \ce{NH2 + CHO -> NH2CHO}
    \label{nh2_cho_ice}
\end{equation}
and the successive hydrogenation of HNCO \citep{raunier2004,garrod2008,jones2011,rimola2018,enrique-romero2019,haupa2019}:
\begin{eqnarray}
    \ce{&H + HNCO &-> NH2CO}\nonumber \\
    \ce{&H + NH2CO &-> NH2CHO}\label{h_hnco},   
\end{eqnarray}
although experimental work by \citet{noble2015} indicates that HNCO cannot be hydrogenated. A third options is the reaction of CN radicals with water ice as proposed in theoretical work by \citet{rimola2018}:
\begin{eqnarray}
    \ce{CN + H2O -> NH2CO}\nonumber \\
    \ce{NH2CO + H -> NH2CHO}\label{cn_h2o_ice}.
\end{eqnarray}

The aforementioned influence of the hydrogen density on the NH$_{2}$CN formation reaction would also be expected to influence NH$_{2}$CHO formation in the ice. Reaction~\ref{nh2_cho_ice} would become less important, instead favoring the hydrogenation of \ce{NH2} to \ce{NH3}:
\begin{equation}
    \ce{NH2 + H -> NH3},
\end{equation}
while at the same time the increased availability of atomic hydrogen could increase the efficiency of Reaction~\ref{h_hnco}, potentially allowing it to dominate. On the other hand, gas phase production of NH$_{2}$CHO can increase under the condition that gas phase NH$_{2}$ and H$_{2}$CO densities increase. This can happen as the gas phase density increases from low-mass to high-mass sources. Finally, top-down reaction pathways can contribute to an increased NH$_{2}$CHO column density. For example, radiation interacting with gas phase CH$_{3}$C(O)NH$_{2}$ can produce the NH$_{2}$CO radical, which in turn can produce NH$_{2}$CHO \citep[e.g.,][]{spall1957}. If radiation fields increase in strength from low-mass to high-mass sources, such a channel can become prominent and result in a lower [NH$_{2}$CN]/[NH$_{2}$CHO] ratio. However, for larger molecules, such as dipeptides, experiments show that photolysis of these species primarily leads to the breaking of the peptide bond itself, instead of its surrounding bonds \citep[][at conditions not relevant to the interstellar medium]{johns1968,neubacher1977}. It is therefore questionable whether top-down chemistry can efficiently produce NH$_{2}$CHO (see also section \ref{sec:HNCO_chem}), but dedicated experiments under interstellar conditions are needed to answer this question. In general, care has to be taken with these chemical explanations, as they will influence formation pathways of, and ratios between, other species as well. Yet, trends like [NH$_{2}$CN]/[NH$_{2}$CHO] are not observed for most other species, such as HNCO and CH$_{3}$C(O)NH$_{2}$, see Fig. \ref{fig:ratios_formamide}.
 
Alternatively, the ice mantle desorption characteristics can explain the observed trend. Thermal desorption depends on the binding energy of the molecule. Recent laboratory studies using temperature programmed desorption investigated the binding energy of NH$_{2}$CHO on graphite and on amorphous water deposited onto graphite \citep{chaabouni2018}. In the latter case, NH$_{2}$CHO remained on the surface until after the water had already desorbed. Those authors found binding energies of 4810 and 5056--6990~K for amorphous water and for formamide, respectively (assuming a desorption-rate prefactor of 10$^{12}$ s$^{-1}$); the binding energy for NH$_{2}$CHO corresponds simply to binding to the graphite left behind following the desorption of the water. The desorption energy of NH$_{2}$CN is not experimentally determined, but chemical models commonly use binding energies of 5556~K for both species, based on the interpolation method described by \citet{garrod2006}. If the true value for NH$_{2}$CN were close to or less than the binding energy of water, or otherwise much less than that of NH$_{2}$CHO, it could influence the [NH$_{2}$CN]/[NH$_{2}$CHO] trend. When a source is sufficiently warm over a large area, such as a HMSFR, most volatile organic species will desorb from the grains. However, a more compact source, such as a single low-mass protostar, will have a decreasing radial temperature profile. Around such sources, species with a low desorption energy desorb over a larger area than those with a high desorption energy. Depending on the spatial resolution of the observations, the beam may simply cover a larger emitting area of NH$_{2}$CN than of NH$_{2}$CHO, thus explaining the decreasing ratio.  To better understand how desorption characteristics influence the [NH$_{2}$CN]/[NH$_{2}$CHO] ratio, laboratory measurements of the NH$_{2}$CN desorption energy are required, but non-thermal desorption processes of NH$_{2}$CN and NH$_{2}$CHO, such as reactive desorption and photodesorption, need to be studied as well. 
 

\subsubsection{HNCO chemistry}
\label{sec:HNCO_chem}

For many years, an abundance correlation between HNCO and NH$_{2}$CHO has been observed toward a large range of interstellar sources, which has led to the belief that there is a direct chemical relationship between these two species \citep[see][for a recent review]{lopez-sepulcre2019}. In Fig. \ref{fig:ratios_formamide}, the [HNCO]/[NH$_{2}$CHO] ratios found toward NGC~6334I are compared with ratios from a select number of other sources. We note that the HNCO column density is determined from its $^{13}$C isotope, multiplied by a $^{12/13}$C ratio of 62. In NGC~6334I~MM1 the [HNCO] / [NH$_{2}$CHO] is slightly above one, while in MM2 it is somewhat higher at a ratio of $\sim$5. Therefore, the ratios in NGC~6334I follows the average interstellar trend and strengthens the probable link between HNCO and NH$_{2}$CHO. Whether this is a direct chemical link \citep[e.g.,][]{raunier2004,haupa2019} or an effect caused by similar chemical responses to physical conditions \citep{quenard2018} cannot be inferred from these ratios. Since HNCO and CH$_{3}$C(O)NH$_{2}$ are found co-spatially toward NGC~6334I, gas phase ion-molecule destruction of CH$_{3}$C(O)NH$_{2}$ may be to some extent responsible for the observed abundances of HNCO, as was suggested by \citet{garrod2008} and \citet{tideswell2010}. In aforementioned works, CH$_{3}$C(O)NH$_{2}$ forms by radical-radical reactions on grain surfaces, together with NH$_{2}$CHO, where after it is released to the gas phase and can be destroyed by ions. This pathway therefore provides an indirect link between HNCO and NH$_{2}$CHO, by way of CH$_{3}$C(O)NH$_{2}$ and other larger amides, assuming that NH$_{2}$CHO and larger amides form in related reactions in ice mantles. This mechanism can help explain the higher [HNCO] / [NH$_{2}$CHO] ratio in MM2, if more ions are present in MM2, for example due to a stronger radiation field, which result in enhanced HNCO production.


\subsubsection{The large amides}

Of the large amides, CH$_{3}$C(O)NH$_{2}$ is detected toward most positions. A relatively small abundance difference is seen between MM1 and MM2, but in general ratios are quite similar throughout the region. This suggests that production of CH$_{3}$C(O)NH$_{2}$ is occurring through a similar mechanism throughout the region. 

Various gas phase formation routes of CH$_{3}$C(O)NH$_{2}$ have been proposed and computationally tested \citep{hollis2006,quan2007,halfen2011,redondo2014}, with the reaction 
\begin{eqnarray}
    \ce{NH2CHO + CH5+ &->& CH3C(O)NH3+ + H2} \nonumber \\
    \ce{CH3C(O)NH3+ + e- &->& CH3C(O)NH2 + H} \label{nh2cho_ch5p}
\end{eqnarray}
considered to be the most efficient. In ice mantles, however, \ce{CH3C(O)NH2} is thought \citep[e.g.,][]{agarwal1985,ligterink2018a} to be primarily produced by Reaction~\ref{ch3_nh2co}:
\begin{equation}
    \ce{CH3 + NH2CO -> CH3C(O)NH2}.
    \label{ch3_nh2co}
\end{equation}
Laboratory simulations of chemical reactions in ice mantles of interstellar dust grains found a [CH$_{3}$C(O)NH$_{2}$]/[NH$_{2}$CHO] ratio of 0.4 upon desorption from the surface \citep{ligterink2018a}, close to the observed average of $\sim$0.15 in NGC~6334I. Ice formation of CH$_{3}$C(O)NH$_{2}$, in particular during the dense cloud stage of star formation, is a plausible mechanism to explain its uniform ratio derived throughout NGC~6334I. While abundances also seem to be similar between NGC~6334I and other sources, it is too early to say if the general interstellar CH$_{3}$C(O)NH$_{2}$ chemistry is the result of the same process(es). A larger sample with confirmed detections and better constraints on the source sizes of the emitting regions toward these objects is required  

Both CH$_{3}$C(O)NH$_{2}$ and CH$_{3}$NHCHO are present at $\sim$10\% with respect to NH$_{2}$CHO toward NGC~6334I. This similarity in abundances supports the conclusion of \citet{belloche2017} that kinetic processes are at the basis of the formation of these species rather than thermal equilibrium. This species can form by energetically processing frozen CH$_{3}$NH$_{2}$:CO mixtures \citep{frigge2018} and thus can be formed on interstellar dust grains. We note that CH$_{3}$NH$_{2}$ is abundantly present toward NGC~6334I \citep{bogelund2019a}. CH$_{3}$NHCHO abundances are similar between NGC~6334I and Sgr B2(N1S), the only other source where this molecule is securely detected \citep{belloche2019}. However, compared to a tentative identification toward Sgr B2(N2), a difference of almost an order of magnitude is seen. At this point it not possible to draw any conclusions on the overall interstellar chemistry of CH$_{3}$NHCHO and more observations of this molecule are required.

The ratios of the tentative detection and upper limits of NH$_{2}$C(O)NH$_{2}$ vary about an order of magnitude within NGC~6334I. It is possible that high temperatures are needed to desorb NH$_{2}$C(O)NH$_{2}$ from dust grains, thus causing this variation. We note that this molecule is tentatively detected toward position MM1-v, which has the highest excitation temperature, and perhaps physical temperature, of 285~K. Compared to secure detections of NH$_{2}$C(O)NH$_{2}$ toward Sgr B2(N1S) and G+0.693 \citep{belloche2019,jimenez-serra2020}, abundances seem to be similar at the few percent level. However, as with the case of CH$_{3}$NHCHO, it is too early to draw conclusions about the formation of NH$_{2}$C(O)NH$_{2}$ at this stage. The tentative identification of NH$_{2}$C(O)NH$_{2}$ warrants future searches for this molecule toward NGC~6334I in order to securely detect this species, better constrain its abundance, and refine and constrain chemical models to elucidate its chemistry.


\subsection{Implications for biomolecule formation}

The detection of CH$_{3}$C(O)NH$_{2}$ and tentative identification of CH$_{3}$NHCHO in NGC~6334I strengthens the case that these molecules can contribute to the formation of biomolecules. It is possible that, in the interstellar medium, these molecules act as precursor species to, or form in parallel with, larger biomolecules. Multiple laboratory experiments show that in UV irradiated simulated ice mantles, consisting of simple molecules such as H$_{2}$O, NH$_{3}$, CH$_{3}$OH, and CH$_{4}$, simple amides and molecules of greater complexity, including biomolecules, form simultaneously \citep[e.g.,][]{agarwal1985,munoz-caro2002,munoz-caro2003,nuevo2010,ligterink2018a,oba2019}. If the formation of amides in the interstellar medium primarily occurs on interstellar dust grains, it is not unlikely that biomolecules are present as well, at least on grain surfaces. Their presence and abundance in the gas phase depends on the local physical conditions and parameter such as binding energy and photodissociation rate. 

Alternatively, amides can also be involved in the earliest stages of biochemistry after delivery to a newly formed planet. Under early Earth conditions and catalyzed by minerals, metals, or clays, NH$_{2}$CHO can form a variety of biomolecules, most notably nucleobases and amino acids \citep[][and references therein]{saladino2012}. Cyanamide and carbamide are known to act as condensing agents in amino acids and nucleotide polymerization reactions \citep{ibanez1971,sakurai1984}. It is not unlikely that CH$_{3}$C(O)NH$_{2}$ and CH$_{3}$NHCHO engage in similar reactions, but laboratory investigations on this topic are limited.   

\begin{figure}[ht!]
\includegraphics[width=0.5\textwidth]{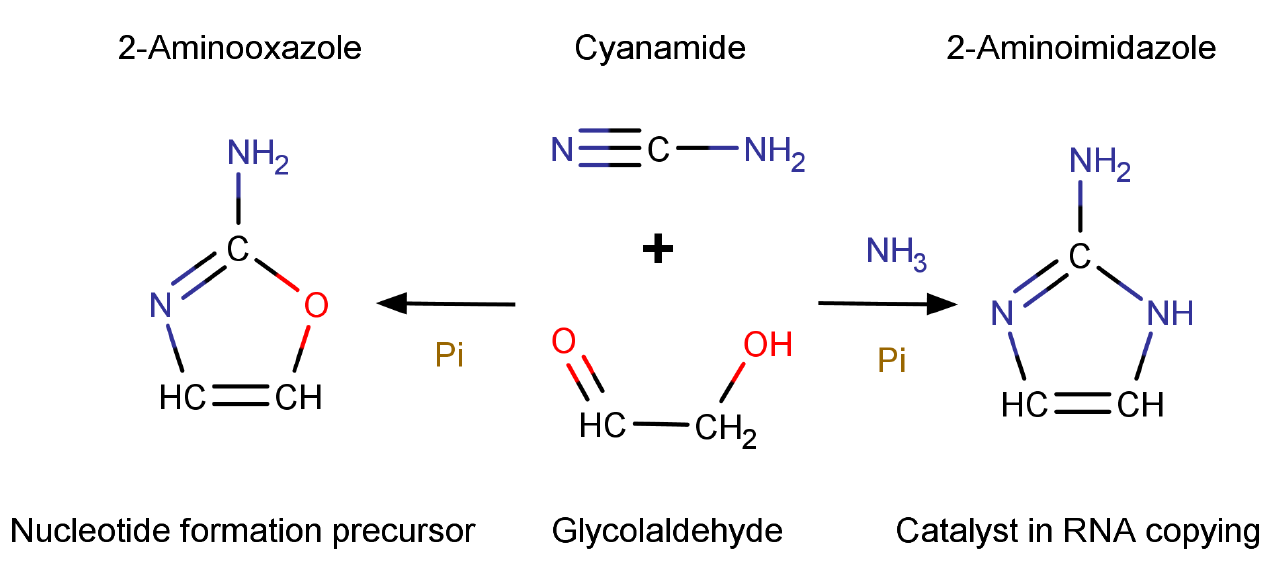}
\caption{Reactions involving the interstellar molecules NH$_{2}$CN and CH$_{2}$(OH)CHO. In a liquid water environment and catalyzed by inorganic phosphates (P$_{i}$), either 2-aminooxazole (left), a nucleotide precursor, or, when NH$_{3}$ is added, 2-aminoimidazole (right), a catalyst for RNA copying, is formed. Figure adopted from \cite{fahrenbach2017}. \label{fig:preb_chem}}
\end{figure}

The cocktail of molecules found toward NGC~6334I, but also other sources such as the sun-like protostar IRAS~16293B, presents more options to directly form biomolecules on a young Earth. This is highlighted by the reaction between NH$_{2}$CN and glycolaldehyde (HC(O)CH$_{2}$OH), resulting in either 2-aminooxazole or 2-aminoimidazole, a precursor to nucleotide formation and a catalyst for RNA copying, respectively  \citep[see Figure \ref{fig:preb_chem},][]{patel2015,fahrenbach2017}. In aforementioned works, both NH$_{2}$CN and HC(O)CH$_{2}$OH are part of $in-situ$ chemistry on a planetary body (Earth in this case) deriving from HCN and involve multiple reaction steps to produce these species. Formation of these two species in space and subsequent delivery to a young Earth directly supplies the $in-situ$ chemical network with these compounds, leading to a more rapid emergence of biomolecules. 
 

\section{Conclusion} \label{sec:conclusion}

In this work, a group of amides is analysed toward the high-mass star-forming region NGC~6334I. Secure detections of NH$_{2}$CN, HNCO, NH$_{2}$CHO, CH$_{3}$C(O)NH$_{2}$, and CH$_{3}$NHCHO are presented along with a tentative identification of NH$_{2}$C(O)NH$_{2}$. Abundances of amides are determined with respect to NH$_{2}$CHO and are generally found to be similar within NGC~6334I, often varying less than a factor of five, hinting that the same chemical pathways and physical conditions are at the basis of this chemistry, despite the large size and physical differences of this HMSFR. It is plausible that production of these amides occurred on surfaces of interstellar dust grains during the dense cloud stage of the star formation process in NGC~6334I. The relatively minor abundance variations between MM1 and MM2 can be the result of slightly different physical conditions, such as higher radiation fields, resulting in the efficient photodestruction of one or more of the amides, either in the ice mantles or gas phase.

The comparison of (tentative) abundances of CH$_{3}$C(O)NH$_{2}$, CH$_{3}$NHCHO, and NH$_{2}$C(O)NH$_{2}$ in NGC~6334I with abundances of these species toward other sources shows that these are quite similar. This hints that similar processes may be responsible for the formation of these species, but due to the limited number of detections and different observing parameters, it is too early to draw any strong conclusion. A different trend is observed for [NH$_{2}$CN] / [NH$_{2}$CHO], for which abundances vary three orders of magnitude over a large number of observed sources. This variation seems to depend on the source type, with high ratios toward low-mass and pre-stellar sources and low ratios toward high-mass sources. The origin of this variation is not easily explained and could depend on different reaction pathways and physical conditions, and is perhaps affected by observational parameters as well. More experimental and theoretical work is needed to confirm this. However, it is likely that NH$_{2}$CN is not part of the same chemistry that produces the other amides. The rich amide chemistry toward NGC~6334I, in combination with other molecules, such as HC(O)CH$_{2}$OH, strengthens the case that interstellar molecules contribute to the rapid emergence of biomolecules on planetary objects.


\acknowledgments
{We thank Dr. E.G. B{\o}gelund for helpful discussions and support in the data analysis. This paper makes use of the following ALMA data: ADS/JAO.ALMA\#2017.1.00717.S, \#2017.1.00661.S, and \#2015.A.00022.T. ALMA is a partnership of ESO (representing its member states), NSF (USA) and NINS (Japan), together with NRC (Canada) and NSC and ASIAA (Taiwan) and KASI (Republic of Korea), in cooperation with the Republic of Chile. The Joint ALMA Observatory is operated by ESO, AUI/NRAO and NAOJ.   The National Radio Astronomy Observatory is a facility of the National Science Foundation operated under cooperative agreement by Associated Universities, Inc.  Support for B.A.M. was provided by NASA through Hubble Fellowship grant \#HST-HF2-51396 awarded by the Space Telescope Science Institute, which is operated by the Association of Universities for Research in Astronomy, Inc., for NASA, under contract NAS5-26555. This research made use of NASA's Astrophysics Data System  Bibliographic  Services,  Astropy,  a community-developed core Python package for Astronomy, and APLpy, an open-source plotting package for Python hosted at http://aplpy.github.com.}


\appendix
\label{sec:appendix}


\section{Chemical structure and naming of amides}
\label{ap:naming}

According to the definition given by the International Union of Pure and Applied Chemistry (IUPAC), amides are oxoacids, which follow the general chemical formula -E(=O)(OH) (E = carbon, sulphur, or other atom that can make this bond), in which the OH group has been replaced by an NH$_{x}$ group. Proper amides are therefore molecules that, for example, follow the structures R$_{1}$-C(=O)N-R$_{2}$R$_{3}$. However, several molecules that do not adhere to this structure are also called amides, such as cyanamide (NH$_{2}$CN). Despite technically not being an amide, cyanamide and the other amides covered in this work do share a distinct chemical structure by which they can be grouped together. 

All amides discussed in this work have structures involving a double (C=O) or triple bonded (C$\equiv$N) group attached to a -NH$_{2}$ group. The molecular orbitals of these fragments are quite different. The double or triple bonded carbon atom and its counterpart have a flat SP2 and SP hybridized molecular orbital, respectively. Normally, a -NH$_{2}$ group will have a tetrahydral SP3 orbital, similar to for example methane (CH$_{4}$). Two molecular fragments that form a bond through a SP2 or SP hybridized orbital will form a rigid structure, but fragments that form a bond through a SP3 orbital can usually freely rotate around the bond axis. However, in the case of an amide, the -NH$_{2}$ group is attached to a carbon atom with SP2 or SP orbital, causing it to take on an SP2 character, forming a so-called resonance hybrid. This effectively means that the (C=X)-NH$_{2}$ bond in an amide takes on a double bond character and becomes a rigid and flat structure. This chemical structure is found in the proper amides (e.g., NH$_{2}$CHO), but also in cyanamide (NH$_{2}$CN), thus making this a distinct feature of this group of molecules.


\section{Spectroscopic linelists}
\label{sec:spec_used}

Spectroscopic linelists used in the work are primarily taken from the CDMS database, but several list from the JPL spectroscopic database and literature sources are used as well. Table \ref{tab:spec_lit} gives an overview of the catalogues and identifiers for each molecule analyzed in this work and the main works these entries are based on. 

\begin{deluxetable}{ccc}[H]
\tablecaption{Linelists and spectroscopic literature \label{tab:spec_lit}}
\tablewidth{0pt}
\tablehead{
\colhead{Molecule} & \colhead{Catalogue} & \colhead{Literature} \\
& \colhead{\& ID} &
}
\startdata
NH$_{2}$CN & JPL & \cite{read1986} \\
& 42003 & \\
HN$^{12}$CO & CDMS & \cite{kukolich1971a} \\
& 43511 & \cite{hocking1975} \\
& & \cite{niedenhoff1995} \\
& & \cite{lapinov2007} \\
HN$^{13}$CO & JPL & \cite{hocking1975} \\
& 44008 & \cite{} \\
NH$_{2}^{12}$CHO & CDMS & \cite{motiyenko2012}\\
& 45512 & \cite{kryvda2009} \\
& & \cite{blanco2006} \\
& & \cite{vorobeva1994} \\
& & \cite{moskienko1991} \\
& & \cite{gardner1980} \\
& & \cite{hirota1974} \\
& & \cite{kukolich1971b} \\
NH$_{2}^{13}$CHO & CDMS & \cite{motiyenko2012}\\
& 46512 & \cite{gardner1980} \\
& & \cite{blanco2006} \\
& & \cite{kryvda2009} \\
CH$_{3}$C(O)NH$_{2}$ & Literature &  \cite{ilyushin2004} \\
CH$_{3}$NHCHO & Literature  & \cite{belloche2017} \\ 
NH$_{2}$C(O)NH$_{2}$ & CDMS & \cite{remijan2014}\\
& 60517 & \cite{brown1975} \\
& & \cite{kasten1986} \\
& & \cite{kretschmer1996} \\
\enddata
\end{deluxetable}


\section{Identified spectral lines}
\label{sec:line_ID}

\startlongtable
\begin{deluxetable*}{lccccc}
\tablecaption{Overview of spectral lines of amides molecules toward NGC~6334I \label{tab:lines_amides}}
\tablewidth{0pt}
\tablehead{
\colhead{Molecule} & \colhead{transition} & \colhead{Frequency$^{\dagger}$} & \colhead{$A_{\rm ij}$} & \colhead{$E_{\rm up}$} \\
\colhead{} & \colhead{$J',K_{\rm a}',K_{\rm c}' - J'',K_{\rm a}'',K_{\rm c}''$} & \colhead{(MHz)} & \colhead{(s$^{-1})$} & \colhead{(K)} 
}
\startdata
NH$_{2}$CN & 14 4 11 - 13 4 10 & 279 479.5918 (0.0066) & 2.05$\times$10$^{-3}$ & 398 \\
& 14 4 10 - 13 4 9 & 279 479.5921 (0.0066) & 2.05$\times$10$^{-3}$ & 398 \\
& 14 0 14 - 13 0 13 & 279 600.4593 (0.0049) & 2.24$\times$10$^{-3}$ & 172 \\
& 14 4 10 - 13 4 9 & 279 763.3778 (0.0056) & 2.12$\times$10$^{-3}$ & 332  \\
& 14 4 11 - 13 4 10 & 279 763.3778 (0.0056) & 2.12$\times$10$^{-3}$ & 332 \\
& 14 0 14 - 13 0 13 & 279 818.8464 (0.0060) & 2.30$\times$10$^{-3}$ & 101  \\
& 14 2 13 - 13 2 12 & 279 820.8787 (0.0037) & 2.26$\times$10$^{-3}$ & 159 \\
& 14 2 12 - 13 2 11 & 279 900.1968 (0.0037) & 2.26$\times$10$^{-3}$ & 159 \\
& 14 3 12 - 13 3 11 & 280 133.2385 (0.0044) & 2.23$\times$10$^{-3}$ & 231 \\
& 14 3 11 - 13 3 10 & 280 137.4694 (0.0044) & 2.23$\times$10$^{-3}$ & 231 \\
& 17 1 17 - 16 1 16 & 337 561.3190 (0.0059) & 4.06$\times$10$^{-3}$ & 160 \\
& 17 2 16 - 16 2 15 & 339 122.8867 (0.0056) & 3.90$\times$10$^{-3}$ & 274 \\
& 17 4 14 - 16 4 13 & 339 347.0172 (0.0078) & 3.80$\times$10$^{-3}$ & 444 \\
& 17 0 17 - 16 0 16 & 339 450.5912 (0.0063) & 4.04$\times$10$^{-3}$ & 218 \\
& 17 2 16 - 16 2 15 & 339 750.7631 (0.0046) & 4.09$\times$10$^{-3}$ & 205 \\
\hline
HN$^{12}$CO & 6 1 6 - 5 1 5 & 131 394.2300 (0.0004) & 2.92$\times$10$^{-5}$ & 65 \\
& 6 4 2 - 5 4 1 & 131 733.5888 (0.0029) & 1.41$\times$10$^{-5}$ & 673 \\
& 6 4 3 - 5 4 2 & 131 733.5888 (0.0029) & 1.41$\times$10$^{-5}$ & 673 & \\
& 6 3 4 - 5 3 3 & 131 799.2971 (0.0024) & 2.07$\times$10$^{-5}$ & 397 \\
& 6 3 3 - 5 3 2 & 131 799.2972 (0.0024) & 2.07$\times$10$^{-5}$ & 397 \\
& 6 2 5 - 5 2 4 & 131 845.8900 (0.0014) & 2.61$\times$10$^{-5}$ & 192 \\
& 6 2 4 - 5 2 3 & 131 846.5998 (0.0014) & 2.61$\times$10$^{-5}$ & 192 \\
& 6 0 6 - 5 0 6 & 131 885.7341 (0.0005) & 3.08$\times$10$^{-5}$ & 22 \\
& 16 1 16 - 15 1 15 & 350 333.0590 (0.0100) & 5.97$\times$10$^{-4}$ & 186 \\
& 16 3 13 - 15 3 12 & 351 416.8123 (0.0052) & 5.31$\times$10$^{-4}$ & 518 \\
& 16 3 14 - 15 3 13 & 351 416.7983 (0.0052) & 5.31$\times$10$^{-4}$ & 518 \\
& 16 2 15 - 15 2 14 & 351 537.7954 (0.0030) & 5.75$\times$10$^{-4}$ & 314 \\
& 16 2 14 - 15 2 13 & 351 551.5731 (0.0032) & 5.75$\times$10$^{-4}$ & 314 \\
& 16 0 16 - 15 0 15 & 351 633.2570 (0.0100) & 6.13$\times$10$^{-4}$ & 143 \\
& 40 1 40 - 39 1 39 & 875 046.1710 (0.0031) & 9.50$\times$10$^{-3}$ & 905 \\
& 40 3 38 - 39 3 37 & 877 808.5933 (0.0111) & 8.68$\times$10$^{-3}$ & 1239 \\
& 40 3 37 - 39 3 36 & 877 809.9771 (0.0112) & 8.68$\times$10$^{-3}$ & 1239 \\
& 40 2 39 - 39 2 38 & 878 062.2368 (0.0063) & 9.25$\times$10$^{-3}$ & 1035 \\
& 40 0 40 - 39 0 39 & 878 137.4392 (0.0029) & 9.72$\times$10$^{-3}$ & 864 \\
& 40 2 38 - 39 2 37 & 878 276.9143 (0.0065) & 9.26$\times$10$^{-3}$ & 1035 \\
& 41 1 41 - 40 1 40 & 896 873.8160 (0.0150) & 1.02$\times$10$^{-2}$ & 948 \\
\hline
HN$^{13}$CO & 6 1 6 - 5 1 5 & 131 397.0955 (0.0084) & 2.96$\times$10$^{-5}$ & 65 \\
& 6 0 6 - 5 0 5 & 131 889.4868 (0.0128) & 3.06$\times$10$^{-5}$ & 22 \\
& 16 1 16 - 15 1 15 & 350 340.3407 (0.0908) & 6.18$\times$10$^{-4}$ & 186 \\
& 16 0 16 - 15 0 15 & 351 642.8746 (0.0763) & 6.27$\times$10$^{-4}$ & 143 \\
\hline
NH$_{2}^{12}$CHO & 6 1 5 - 5 1 4 & 131 617.9025 (0.0005) & 1.56$\times$10$^{-4}$ & 25 \\
& 25 2 23 - 25 1 24 & 280 689.4724 (0.0009) & 5.21$\times$10$^{-5}$ & 351 \\
& 12 5 8 - 13 4 9 & 281 381.6715 (0.0012) & 1.04$\times$10$^{-5}$ & 154 \\
& 15 2 14 - 15 1 15 & 281 934.6250 (0.0008) & 3.40$\times$10$^{-5}$ & 134 \\
& 14 3 11 - 14 2 12 & 282 822.5932 (0.0009) & 5.44$\times$10$^{-5}$ & 134 \\
& 15 2 14 - 15 0 15 & 294 753.3491 (0.0008) & 1.07$\times$10$^{-5}$ & 134 \\
& 14 2 13 - 13 2 12 & 294 776.9896 (0.0009) & 1.84$\times$10$^{-3}$ & 118 \\
& 17 3 15 - 17 2 16 & 336 733.0134 (0.0009) & 8.20$\times$10$^{-5}$ & 183 \\
& 16 13 3 - 15 13 2 & 339 746.2586 (0.0023) & 9.84$\times$10$^{-4}$ & 640 \\
& 16 13 4 - 15 13 3 & 339 746.2586 (0.0023) & 9.84$\times$10$^{-4}$ & 640 \\
& 16 7 10 - 15 7 9 & 339 779.5370 (0.0300) & 2.34$\times$10$^{-3}$ & 284 \\
& 16 7 9 - 15 7 8 & 339 779.5370 (0.0300) & 2.34$\times$10$^{-3}$ & 284 \\
& 16 2 14 - 15 2 13 & 349 478.2047 (0.0009) & 3.10$\times$10$^{-3}$ & 153 \\
& 9 2 8 - 8 1 7 & 349 634.0304 (0.0012) & 6.22$\times$10$^{-5}$ & 58 \\ 
& 41 7 34 - 40 7 33 & 874 836.2150 (0.0500) & 4.89$\times$10$^{-2}$ & 1025 \\
& 42 2 40 - 41 2 39 & 875 818.8790 (0.0500) & 5.01$\times$10$^{-2}$ & 940 \\
& 15 5 10 - 14 4 11 & 876 148.3790 (0.0500) & 1.11$\times$10$^{-3}$ & 197 \\
& 41 5 37 - 40 5 36 & 876 252.7700 (0.0500) & 4.98$\times$10$^{-2}$ & 956 \\
& 41 6 36 - 40 6 35 & 876 493.3340 (0.0500) & 4.95$\times$10$^{-2}$ & 988 \\
& 41 6 35 - 40 6 34 & 878 706.9000 (0.0500) & 4.99$\times$10$^{-2}$ & 988 \\
& 22 3 19 - 21 2 20 & 879 169.0967 (0.0022) & 3.84$\times$10$^{-4}$ & 287 \\
& 42 13 29 - 41 13 28 & 891 600.4990 (0.0500) & 4.82$\times$10$^{-2}$ & 1422 \\
& 42 13 30 - 41 13 29 & 891 600.4990 (0.0500) & 4.82$\times$10$^{-2}$ & 1422 \\
& 43 3 41 - 42 3 40 & 893 045.4010 (0.0500) & 5.32$\times$10$^{-2}$ & 983 \\
& 45 0 45 - 44 1 44 & 893 077.5610 (0.0500) & 2.76$\times$10$^{-3}$ & 997 \\
& 45 1 45 - 44 1 44 & 893 083.8870 (0.0500) & 5.35$\times$10$^{-2}$ & 997 \\
& 45 0 45 - 44 0 44 & 893 085.9940 (0.0500) & 5.35$\times$10$^{-2}$ & 997 \\
& 45 1 45 - 44 0 44 & 893 092.3230 (0.0500) & 2.76$\times$10$^{-3}$ & 997 \\
& 44 2 43 - 43 2 42 & 893 184.3320 (0.0500) & 5.34$\times$10$^{-2}$ & 992 \\
& 42 9 34 - 41 9 33 & 893 415.4200 (0.0500) & 5.12$\times$10$^{-2}$ & 1162 \\
& 42 9 33 - 41 9 32 & 893 415.4200 (0.0500) & 5.12$\times$10$^{-2}$ & 1162 \\
& 42 8 35 - 41 8 34 & 894 533.0350 (0.0500) & 5.19$\times$10$^{-2}$ & 1112 \\
& 42 8 34 - 41 8 33 & 894 550.9190 (0.0500) & 5.19$\times$10$^{-2}$ & 1112 \\
& 43 2 41 - 42 2 40 & 895 080.2490 (0.0500) & 5.35$\times$10$^{-2}$ & 983 \\
\hline
NH$_{2}^{13}$CHO & 6 1 5 - 5 1 4 & 131 495.8540 (0.0100) & 1.55$\times$10$^{-4}$ & 25 \\
& 13 1 12 - 12 1 11 & 282 107.1134 (0.0026) & 1.63$\times$10$^{-3}$ & 98 \\
& 14 1 14 - 13 1 13 & 282 693.5048 (0.0027) & 1.65$\times$10$^{-3}$ & 105 \\
& 16 8 8 - 15 8 7 & 339 213.4944 (0.0029) & 2.16$\times$10$^{-3}$ & 324 \\
& 16 8 9 - 15 8 8 & 339 213.4944 (0.0029) & 2.16$\times$10$^{-3}$ & 324 \\
& 16 2 14 - 15 2 13 & 349 308.8534 (0.0030) & 3.10$\times$10$^{-3}$ & 153 \\
\hline
CH$_{3}$C(O)NH$_{2}$ & 23 10 14 - 22 9 13 & 351 083.4533 (0.0458) & 2.12$\times$10$^{-3}$ & 235 \\
 & 22 10 12 - 21 11 11 & 351 097.8247 (0.0412) & 1.89$\times$10$^{-3}$ & 225 \\
 & 22 11 12 - 21 10 11 & 351 097.8510 (0.0412) & 1.89$\times$10$^{-3}$ & 225 \\
 & 24 8 16 - 23 9 15 & 351 134.2229 (0.0499) & 2.32$\times$10$^{-3}$ & 244 \\
 & 24 9 16 - 23 8 15 & 351 134.2229 (0.0499) & 2.32$\times$10$^{-3}$ & 244 \\
 & 25 7 18 - 24 8 17 & 351 214.4618 (0.0534) & 2.51$\times$10$^{-3}$ & 253 \\
 & 25 8 18 - 24 7 17 & 351 214.4618 (0.0534) & 2.51$\times$10$^{-3}$ & 253 \\
 & 26 6 20 - 25 7 19 & 351 309.0658 (0.0563) & 2.69$\times$10$^{-3}$ & 260 \\
 & 26 7 20 - 25 6 19 & 351 309.0658 (0.0563) & 2.69$\times$10$^{-3}$ & 260 \\
 & 29 3 26 - 28 4 25 & 351 621.6370 (0.0657) & 3.18$\times$10$^{-3}$ & 276 \\
 & 29 4 26 - 28 3 25 & 351 621.6370 (0.0657) & 3.18$\times$10$^{-3}$ & 276 \\
 & 31 1 30 - 30 2 29 & 351 830.6986 (0.0819) & 3.47$\times$10$^{-3}$ & 282 \\
 & 31 2 30 - 30 1 29 & 351 830.6986 (0.0819) & 3.47$\times$10$^{-3}$ & 282 \\
\hline
CH$_{3}$NHCHO & 11 3 8 - 10 3 7$^{\ddagger}$ & 130 190.7247 (0.0009) & 9.21$\times$10$^{-5}$ & 43 \\
 & 12 2 11 - 11 2 10 & 130 874.1508 (0.0010) & 9.75$\times$10$^{-5}$ & 44 \\
 & 11 2 9 - 10 2 8$^{\ddagger}$ & 131 427.7286 (0.0009) & 9.86$\times$10$^{-5}$ & 40 \\
 & 13 8 5 - 12 8 4 & 144 416.877 (0.0014) & 8.97$\times$10$^{-5}$ & 97 \\
 & 9 7 2 - 8 6 2 & 279 381.6637 (0.0035) & 5.18$\times$10$^{-4}$ & 62 \\
 & 25 12 13 -24 12 12$^{\ddagger}$ & 280 384.9819 (0.0041) & 8.20$\times$10$^{-4}$ & 274 \\
 & 25 14 11 - 24 14 11$^{\ddagger}$ & 280 716.5012 (0.0202) & 6.82$\times$10$^{-4}$ & 308 \\
 & 25 10 15 - 24 10 14 & 280 743.3675 (0.0049) & 8.94$\times$10$^{-4}$ & 245 \\
 & 24 5 19 - 23 5 18$^{\ddagger}$ & 280 994.8045 (0.0021) & 9.96$\times$10$^{-4}$ & 186 \\
 & 25 13 13 - 24 13 12 & 281 110.7375 (0.0045) & 7.76$\times$10$^{-4}$ & 291 \\
 & 25 13 12 - 24 13 11 & 281 113.0259 (0.0046) & 7.81$\times$10$^{-4}$ & 291 \\
 & 25 11 14 - 24 11 13 & 282 584.3703 (0.0088) & 8.55$\times$10$^{-4}$ & 259 \\
 & 26 9 17 - 25 9 16 & 291 978.5554 (0.0027) & 1.06$\times$10$^{-3}$ & 247 \\
 & 26 12 14 - 25 12 13 & 292 546.3434 (0.0055) & 9.46$\times$10$^{-4}$ & 288 \\
 & 26 5 22 - 25 5 21 & 292 648.065 (0.0023) & 1.13$\times$10$^{-3}$ & 211 \\
 & 26 15 11 - 25 15 10$^{\ddagger}$ & 292 847.7083 (0.0108) & 8.05$\times$10$^{-4}$ & 341 \\
 & 25 5 20 - 24 5 19 & 293 006.2032 (0.0023) & 1.14$\times$10$^{-3}$ & 200 \\
 & 8 8 0 - 7 7 0 & 294 241.2632 (0.0049) & 7.64$\times$10$^{-4}$ & 67 \\
 & 34 0 34 - 33 0 33 & 337 907.4314 (0.0048) & 1.30$\times$10$^{-3}$ & 291 \\
 & 34 1 34 - 33 1 33 & 337 907.4314 (0.0048) & 1.30$\times$10$^{-3}$ & 291 \\
 & 34 1 34 - 33 0 33 & 337 907.4314 (0.0048) & 1.30$\times$10$^{-3}$ & 291 \\
 & 34 0 34 - 33 1 33 & 337 907.4314 (0.0048) & 1.30$\times$10$^{-3}$ & 291 \\
 & 30 10 20 - 29 10 19$^{\ddagger}$ & 337 993.2492 (0.0041) & 1.61$\times$10$^{-3}$ & 321 \\
 & 30 4 26 - 29 4 25 & 338 606.6396 (0.0028) & 1.74$\times$10$^{-3}$ & 272 \\
 & 29 6 23 - 28 6 22 & 348 416.9795 (0.0029) & 1.89$\times$10$^{-3}$ & 268 \\
 & 31 4 27 - 30 4 26 & 348 417.6722 (0.009) & 1.39$\times$10$^{-3}$ & 288 \\
 & 30 5 25 - 29 5 24 & 348 837.7283 (0.0030) & 1.94$\times$10$^{-3}$ & 279 \\ 
 & 31 5 27 - 30 4 26$^{\ddagger}$ & 350 710.9037 (0.0031) & 7.43$\times$10$^{-4}$ & 288 \\
\hline
NH$_{2}$C(O)NH$_{2}$ & 25 3 23 - 24 2 22 & 279 472.8671 (0.0468) & 2.55$\times$10$^{-4}$ & 195 \\
& 25 2 23 - 24 2 22 & 279 472.8671 (0.0468) & 1.25$\times$10$^{-3}$ & 195 \\
& 25 3 23 - 24 3 22 & 279 472.8671 (0.0468) & 1.25$\times$10$^{-3}$ & 195 \\
& 25 2 23 - 24 3 22 & 279 472.8671 (0.0468) & 2.55$\times$10$^{-4}$ & 195 \\
& 24 5 20 - 23 4 19 & 282 190.8684 (0.0399) & 1.20$\times$10$^{-3}$ & 199 \\
& 24 5 20 - 23 5 19 & 282 190.8684 (0.0399) & 2.16$\times$10$^{-4}$ & 199 \\
& 24 4 20 - 23 4 19 & 282 190.8684 (0.0399) & 2.16$\times$10$^{-4}$ & 199 \\
& 24 4 20 - 23 5 19 & 282 190.8684 (0.0399) & 1.20$\times$10$^{-3}$ & 199 \\
& 25 4 21 - 24 4 20 & 290 862.3844 (0.0473) & 1.48$\times$10$^{-3}$ & 213 \\
& 25 4 21 - 24 5 20 & 290 862.3844 (0.0473) & 9.83$\times$10$^{-5}$ & 213 \\
& 25 5 21 - 24 5 20 & 290 862.3844 (0.0473) & 9.83$\times$10$^{-5}$ & 213 \\
& 25 5 21 - 24 4 20 & 290 862.3844 (0.0473) & 1.48$\times$10$^{-3}$ & 213 \\
& 22 7 16 - 21 6 15 & 293 236.5715 (0.0337) & 1.18$\times$10$^{-3}$ & 186 \\
& 22 6 16 - 21 7 15 & 293 236.5646 (0.0337) & 1.18$\times$10$^{-3}$ & 186 \\
& 21 7 14 - 20 8 13 & 293 292.5803 (0.0297) & 1.05$\times$10$^{-3}$ & 179 \\
& 21 8 14 - 20 7 13 & 293 293.5513 (0.0297) & 1.05$\times$10$^{-3}$ & 179 \\
& 24 5 19 - 23 5 18 & 293 642.0215 (0.0385) & 1.01$\times$10$^{-4}$ & 208 \\
& 24 6 19 - 23 5 18 & 293 642.0215 (0.0385) & 1.39$\times$10$^{-3}$ & 208 \\
& 24 5 19 - 23 6 18 & 293 642.0215 (0.0385) & 1.39$\times$10$^{-3}$ & 208 \\
& 24 6 19 - 23 6 18 & 293 642.0215 (0.0385) & 1.01$\times$10$^{-4}$ & 208 \\
\enddata
\tablecomments{$^{\dagger}$The uncertainty of each line frequency is given in the parentheses. $^{\ddagger}$CH$_{3}$NHCHO lines that are present on locations where the baseline dips.}
\end{deluxetable*}


\section{Spectral features toward MM1-i and MM2-i}
\label{sec:mm1-2_spec_feat}

\begin{figure*}[ht!]
\includegraphics[width=\textwidth]{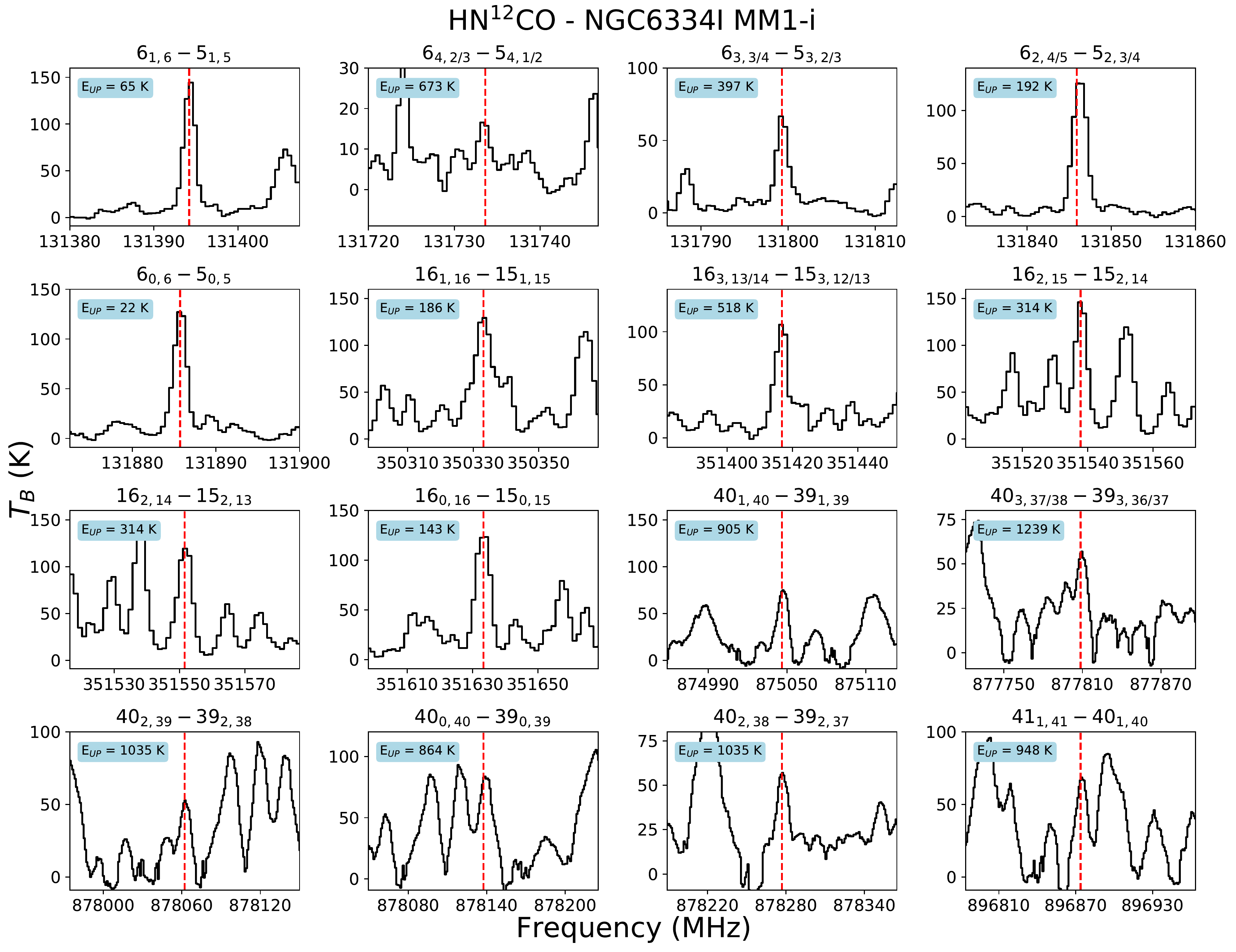}
\caption{Detected transitions of HN$^{12}$CO in the spectrum toward position MM1-i. The observed spectrum is plotted in black and the rest frequency center of the transition is indicated by the red dotted line. Because these lines are optically thick, they are not fitted with a synthetic spectrum. The upper state energy of each transition is given in the top left corner. \label{fig:12-hnco_mm1-i}}
\end{figure*}

\newpage

\begin{figure*}[ht!]
\includegraphics[width=\textwidth]{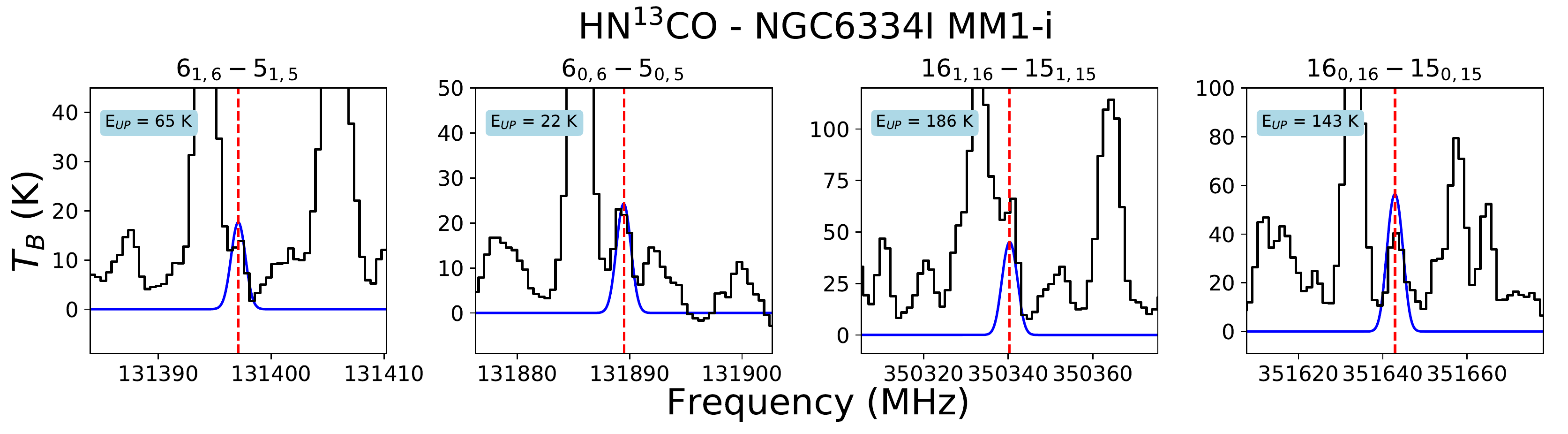}
\caption{Detected transitions of HN$^{13}$CO in the spectrum toward position MM1-i. The observed spectrum is plotted in black, with the synthetic spectrum overplotted in blue and the rest frequency center of the transition is indicated by the red dotted line. The upper state energy of each transition is given in the top left corner. \label{fig:13-hnco_mm1-i}}
\end{figure*}

\newpage

\begin{figure*}[ht!]
\includegraphics[width=\textwidth]{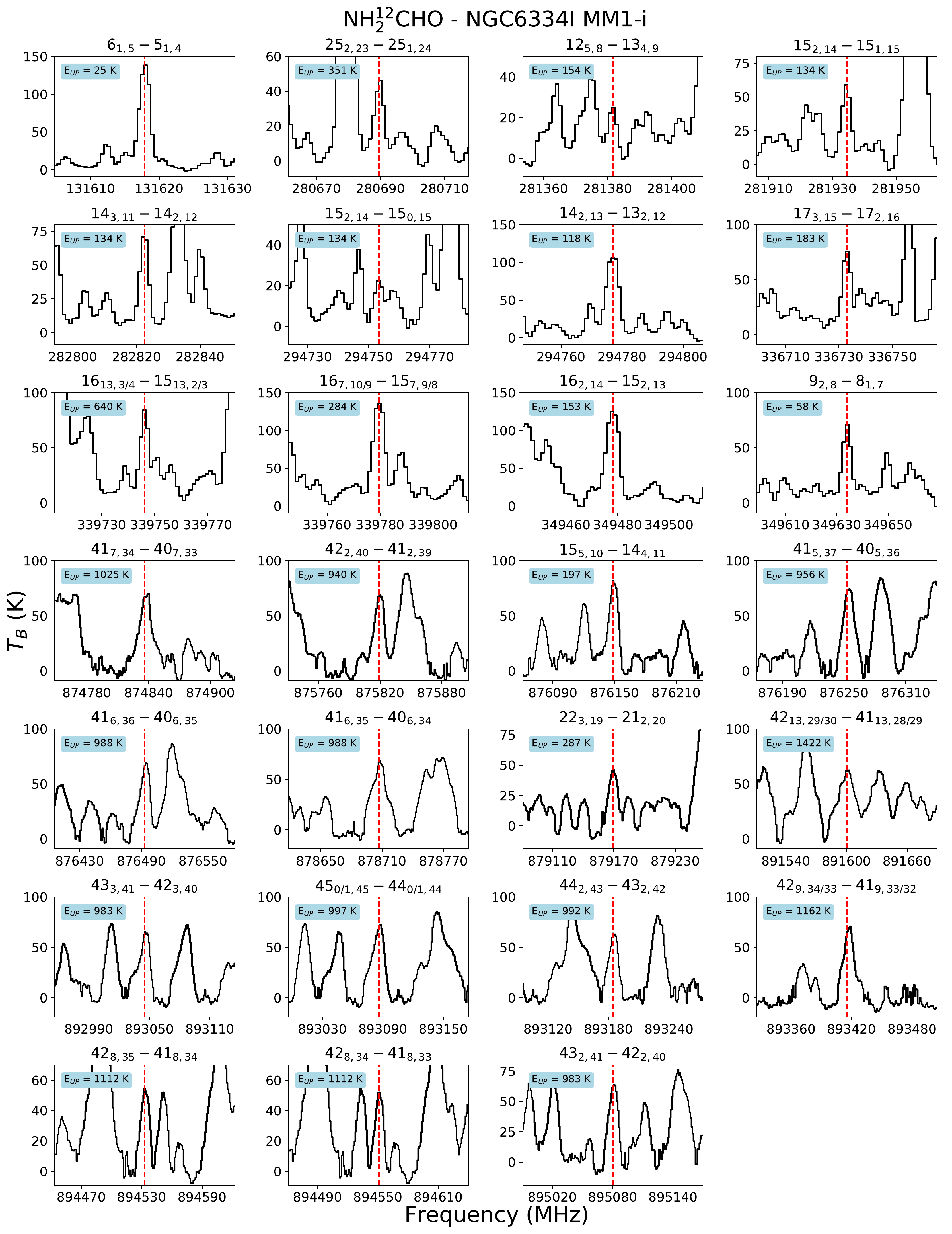}
\caption{Detected transitions of NH$_{2}^{12}$CHO in the spectrum toward position MM1-i. The observed spectrum is plotted in black and the rest frequency center of the transition is indicated by the red dotted line. Because these lines are optically thick, they are not fitted with a synthetic spectrum. The upper state energy of each transition is given in the top left corner. \label{fig:12-nh2cho_mm1-i}}
\end{figure*}

\newpage

\begin{figure*}[ht!]
\epsscale{1.2}
\includegraphics[width=\textwidth]{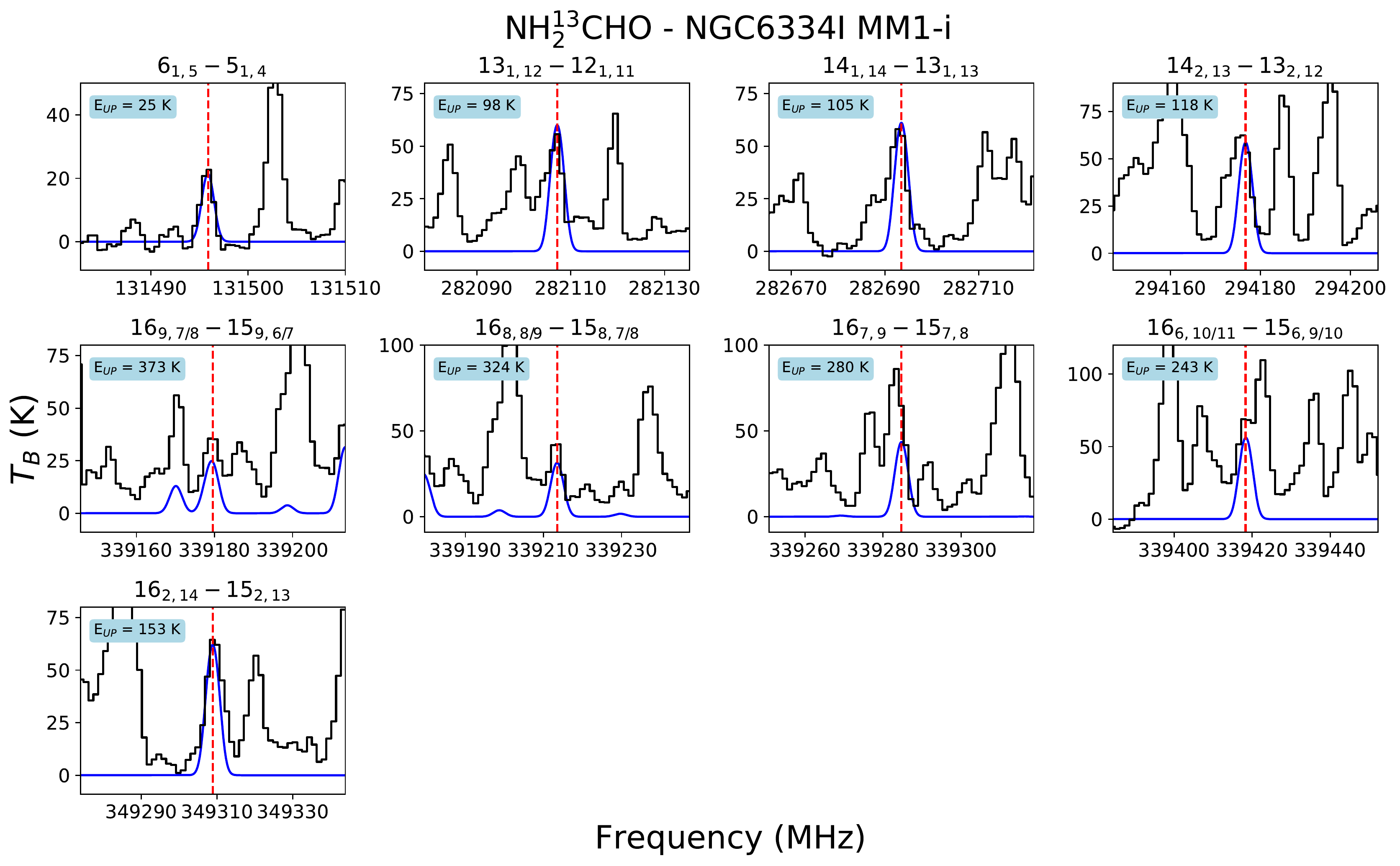}
\caption{Detected transitions of NH$_{2}^{13}$CHO in the spectrum toward position MM1-i. The observed spectrum is plotted in black, with the synthetic spectrum overplotted in blue and the rest frequency center of the transition is indicated by the red dotted line. The upper state energy of each transition is given in the top left corner. \label{fig:13-nh2cho_mm1-i}}
\end{figure*}

\newpage

\begin{figure*}[ht!]
\includegraphics[width=\textwidth]{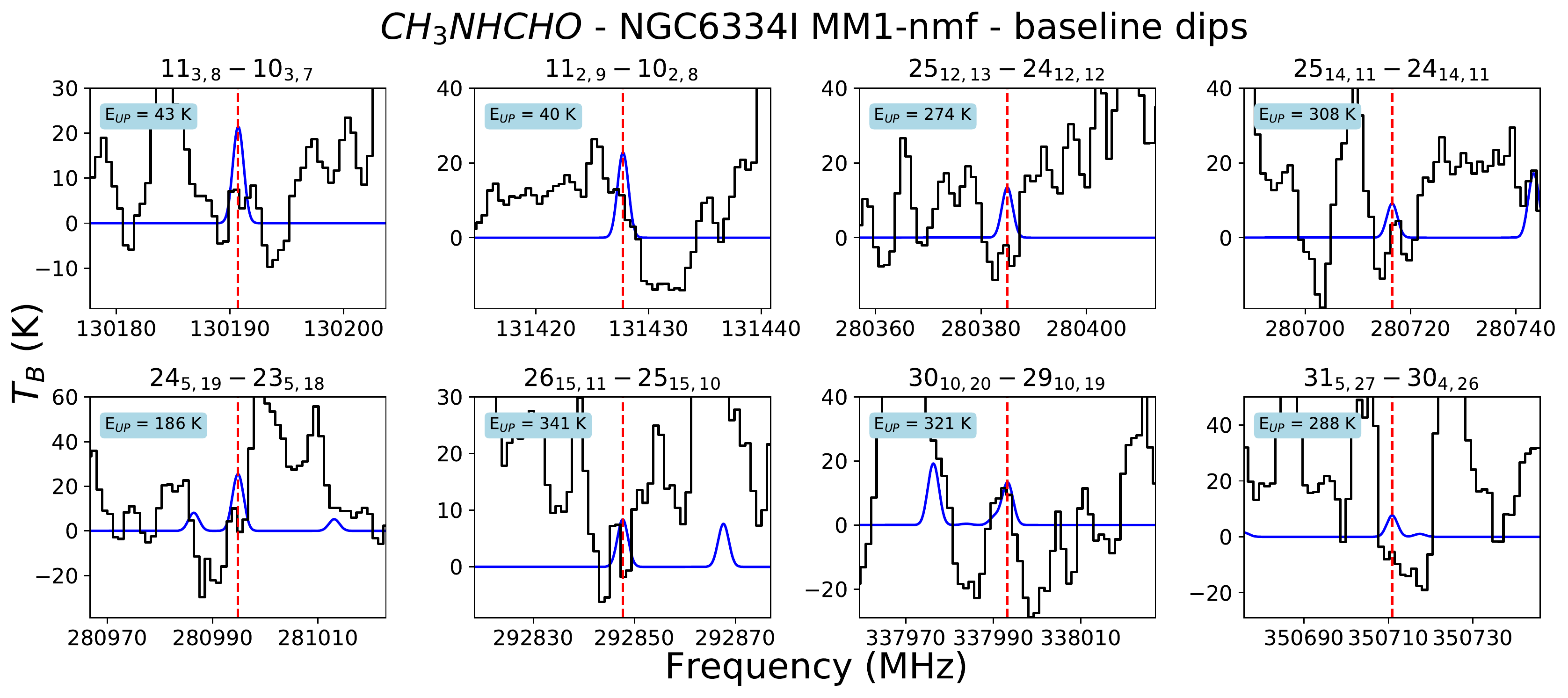}
\caption{Transitions of CH$_{3}$NHCHO that are missing in the spectrum toward position MM1-nmf due to baseline dips. The observed spectrum is plotted in black, with the synthetic spectrum overplotted in blue and the rest frequency center of the transition is indicated by the red dotted line. The upper state energy of each transition is given in the top left corner. \label{fig:nmf_mm1-nmf_dips}}
\end{figure*}

\newpage

\begin{figure*}[ht!]
\includegraphics[width=\textwidth]{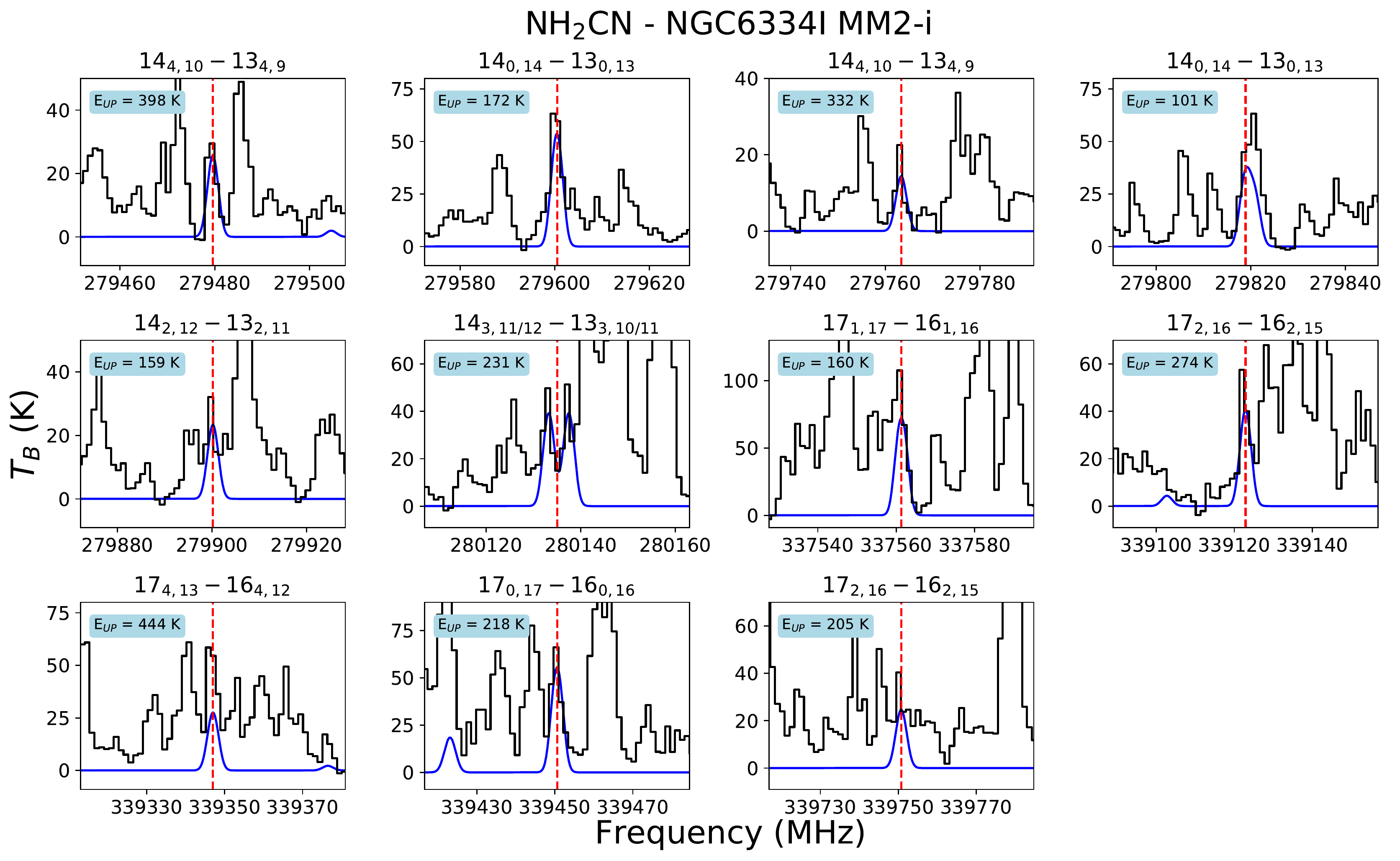}
\caption{Detected transitions of NH$_{2}$CN in the spectrum toward position MM2-i. The observed spectrum is plotted in black, with the synthetic spectrum overplotted in blue and the rest frequency center of the transition is indicated by the red dotted line. The upper state energy of each transition is given in the top left corner. \label{fig:cyanamide_mm2-i}}
\end{figure*}

\newpage

\begin{figure*}[ht!]
\includegraphics[width=\textwidth]{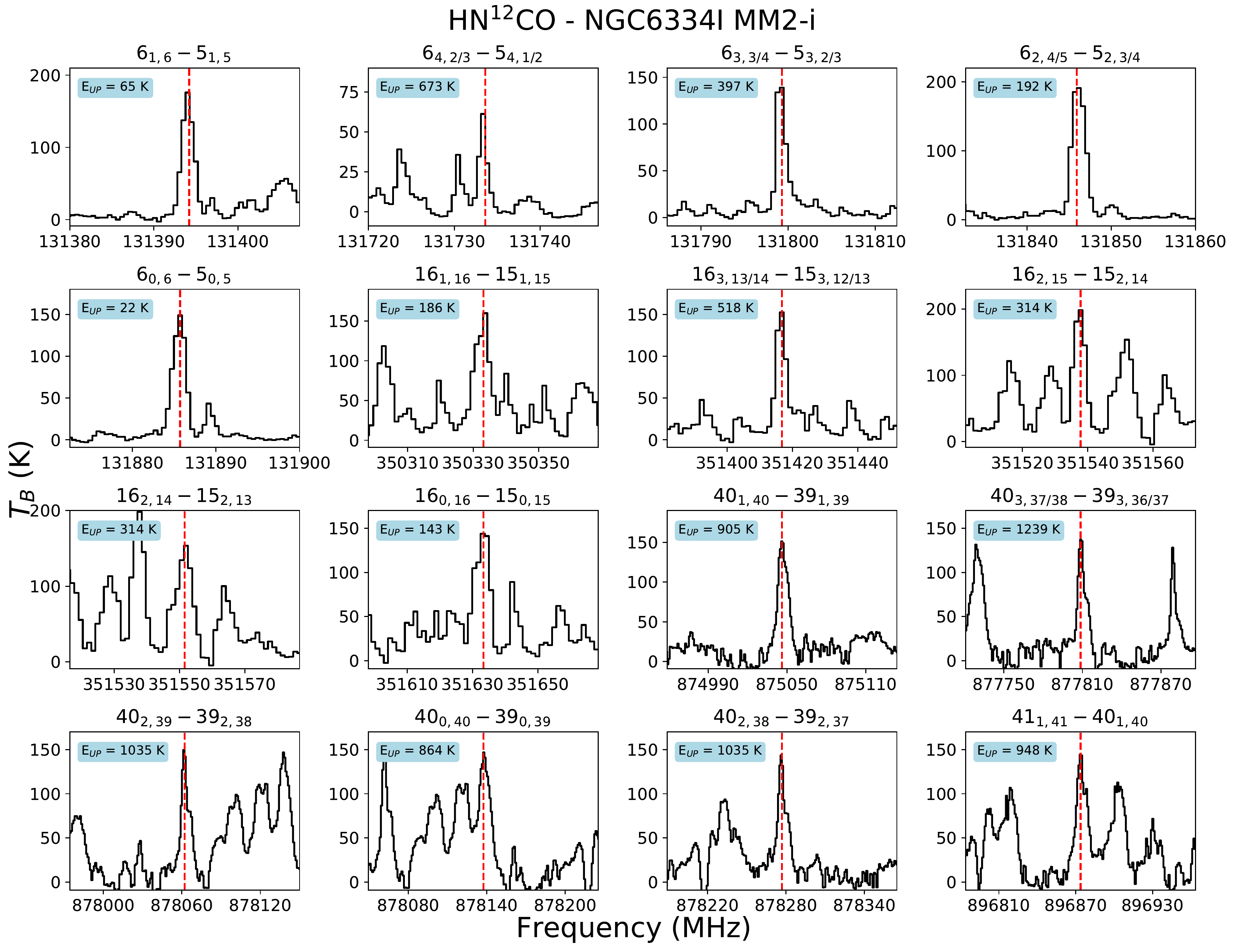}
\caption{Detected transitions of HN$^{12}$CO in the spectrum toward position MM2-i. The observed spectrum is plotted in black and the rest frequency center of the transition is indicated by the red dotted line. Because these lines are optically thick, they are not fitted with a synthetic spectrum. The upper state energy of each transition is given in the top left corner. \label{fig:12-hnco_mm2-i}}
\end{figure*}

\newpage

\begin{figure*}[ht!]
\includegraphics[width=\textwidth]{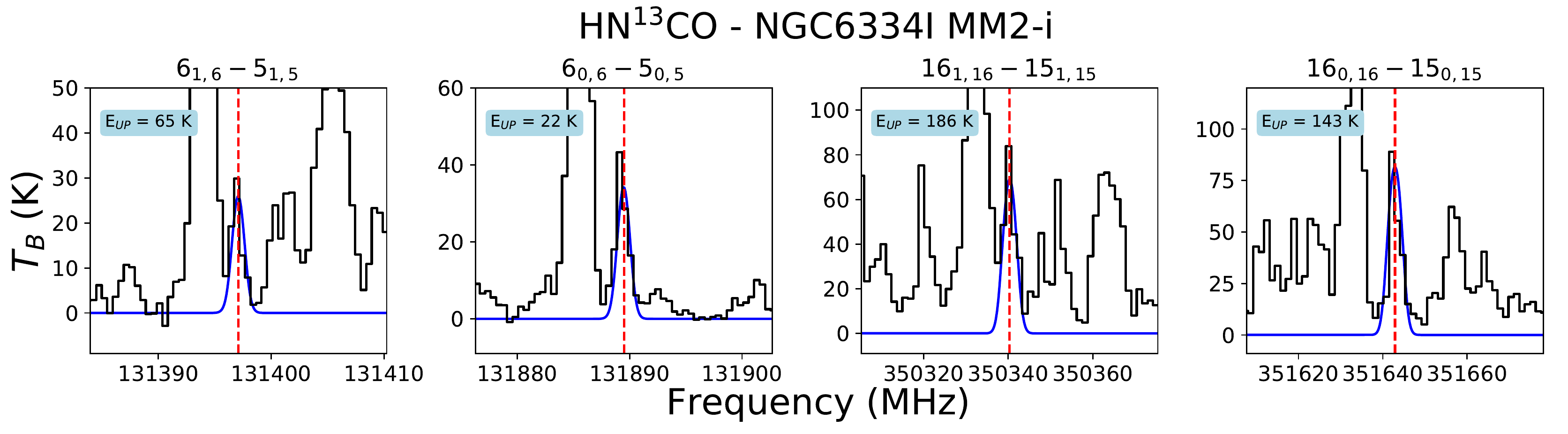}
\caption{Detected transitions of HN$^{13}$CO in the spectrum toward position MM2-i. The observed spectrum is plotted in black, with the synthetic spectrum overplotted in blue and the rest frequency center of the transition is indicated by the red dotted line. The upper state energy of each transition is given in the top left corner. \label{fig:13-hnco_mm2-i}}
\end{figure*}

\newpage

\begin{figure*}[ht!]
\includegraphics[width=\textwidth]{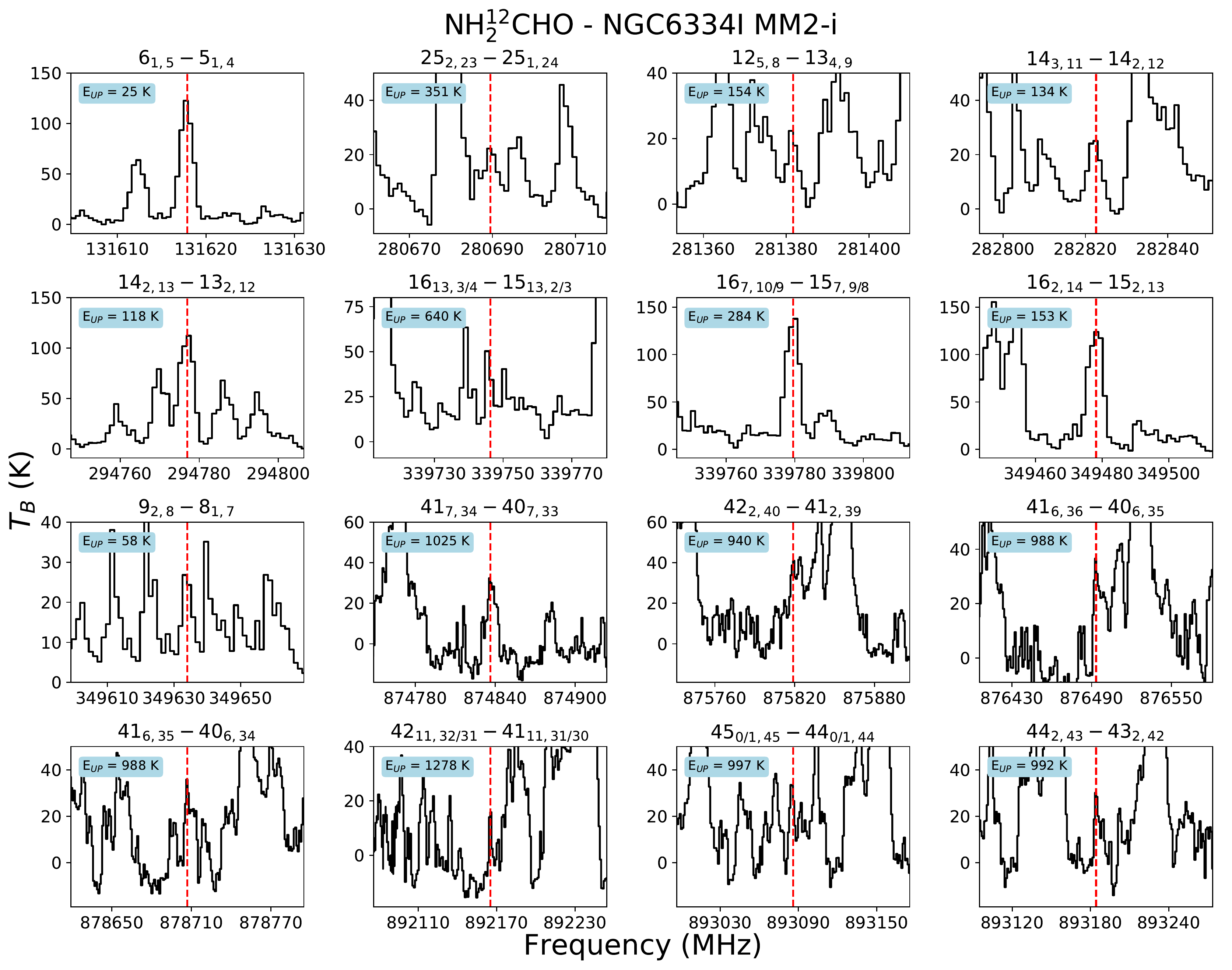}
\caption{Detected transitions of NH$_{2}^{12}$CHO in the spectrum toward position MM2-i. The observed spectrum is plotted in black the rest frequency center of the transition is indicated by the red dotted line. Because these lines are optically thick, they are not fitted with a synthetic spectrum. The upper state energy of each transition is given in the top left corner. \label{fig:12-nh2cho_mm2-i}}
\end{figure*}

\newpage

\begin{figure*}[ht!]
\includegraphics[width=\textwidth]{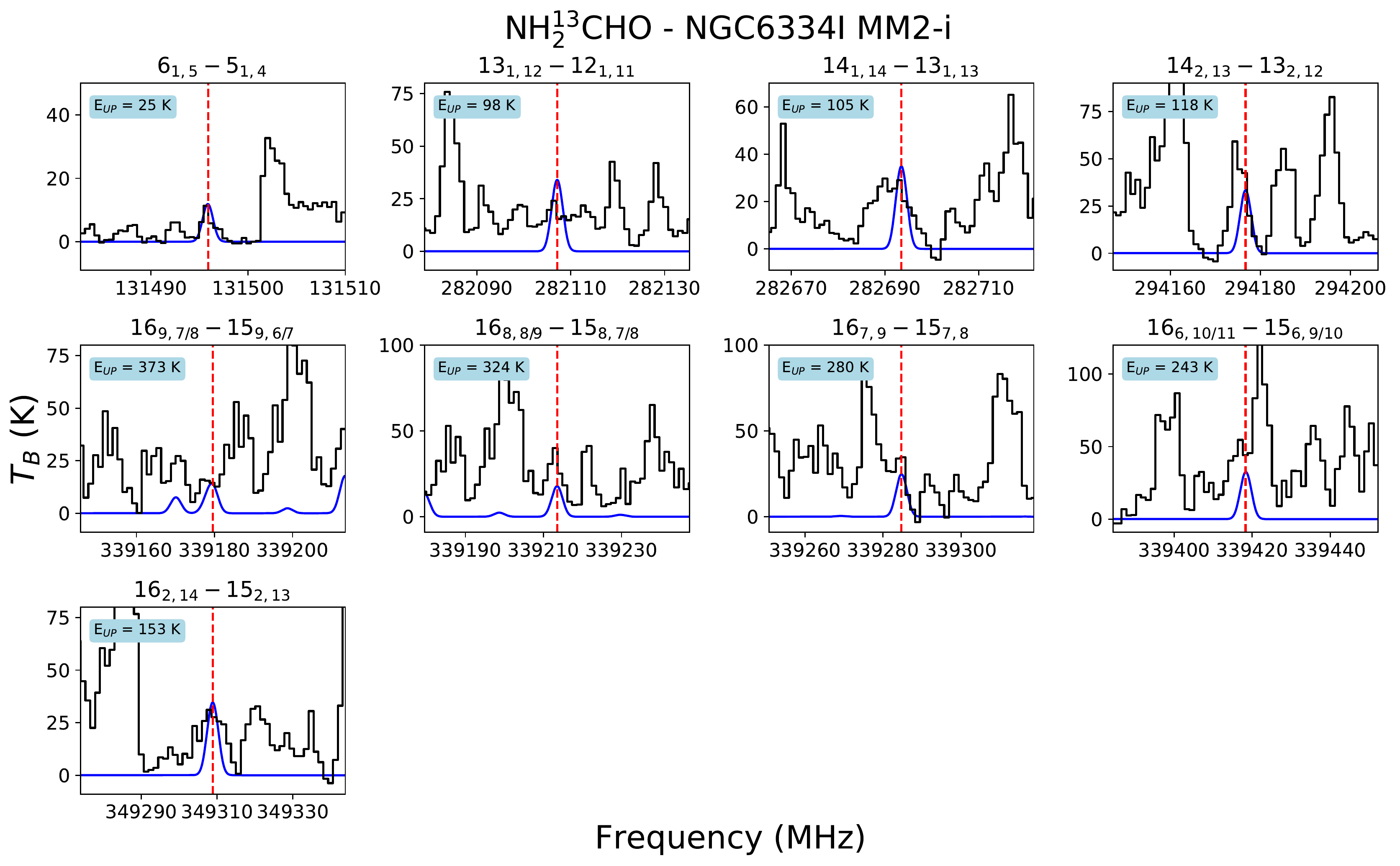}
\caption{Detected transitions of NH$_{2}^{13}$CHO in the spectrum toward position MM2-i. The observed spectrum is plotted in black, with the synthetic spectrum overplotted in blue and the rest frequency center of the transition is indicated by the red dotted line. The upper state energy of each transition is given in the top left corner. \label{fig:13-nh2cho_mm2-i}}
\end{figure*}

\newpage

\begin{figure*}[ht!]
\includegraphics[width=\textwidth]{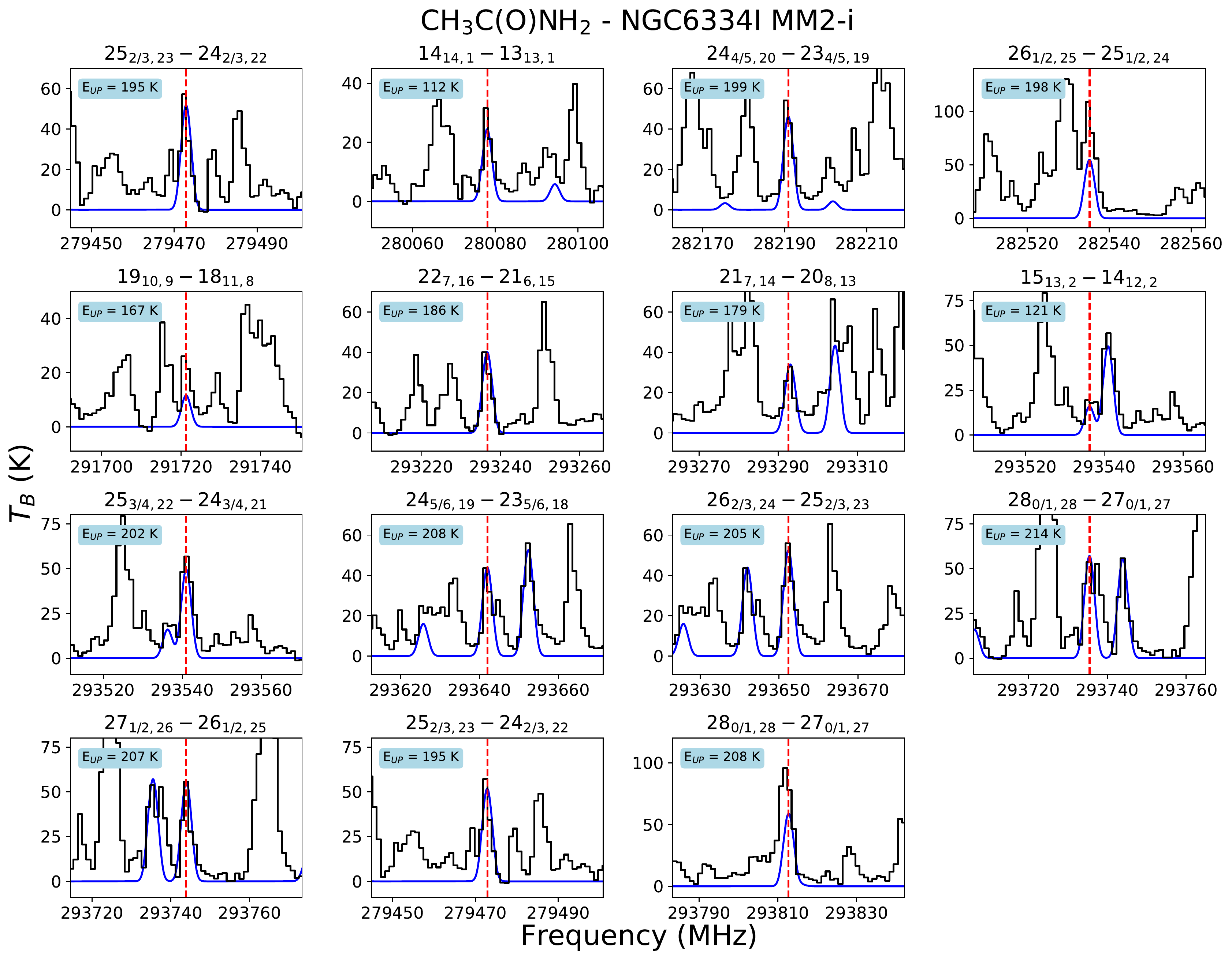}
\caption{Detected transitions of CH$_{3}$C(O)NH$_{2}$ in the spectrum toward position MM2-i. The observed spectrum is plotted in black, with the synthetic spectrum overplotted in blue and the rest frequency center of the transition is indicated by the red dotted line. The upper state energy of each transition is given in the top left corner. \label{fig:acetamide_mm2-i}}
\end{figure*}


\end{document}